\documentclass[twocolumn,tighten,times]{aastex62}
\pdfoutput=1 
\usepackage{amsmath}
\usepackage[T1]{fontenc}
\usepackage{apjfonts} 
\usepackage{anyfontsize}
\usepackage{amssymb}
\usepackage{sidecap}
\usepackage{natbib}
\usepackage{graphicx}

\newcommand{\Kepler}{\textit{Kepler}}
\newcommand{\tess}{\textit{TESS}}
\newcommand{\kepler}{\textit{Kepler}}
\newcommand{\redpen}{}
\newcommand\blfootnote[1]{%
  \begingroup
  \renewcommand\thefootnote{}\footnote{#1}%
  \addtocounter{footnote}{-1}%
  \endgroup
}

\newcommand{\JHU}{Department of Physics and Astronomy, The Johns Hopkins University, Baltimore, MD 21218, USA}
\newcommand{\JHUCS}{the Department of Computer Science, Johns Hopkins University, Baltimore, MD 21218, USA}
\newcommand{\STScI}{Space Telescope Science Institute, Baltimore, MD 21218, USA}
\newcommand{\TCD}{School of Physics, Trinity College Dublin, The University of Dublin, Dublin 2, Ireland}
\newcommand{\UIUC}{Department of Astronomy, University of Illinois at Urbana-Champaign, 1002 W. Green St., IL 61801, USA}
\newcommand{\UIUCCAS}{Center for Astrophysical Surveys, National Center for Supercomputing Applications, Urbana, IL 61801, USA}
\newcommand{\PSU}{Department of Astronomy \& Astrophysics, The Pennsylvania State University, University Park, PA 16802, USA}
\newcommand{\ICDS}{Institute for Computational \& Data Sciences, The Pennsylvania State University, University Park, PA 16802, USA}
\newcommand{\IGC}{Institute for Gravitation and the Cosmos, The Pennsylvania State University, University Park, PA 16802, USA}
\newcommand{\UCSC}{Department of Astronomy and Astrophysics, University of California, Santa Cruz, CA 95064, USA}
\newcommand{\konkoly}{Konkoly Observatory, CSFK, Konkoly-Thege M. ut 15-17, Budapest, 1121 Hungary}
\newcommand{\ELTEphysics}{ELTE E\"otv\"os Lor\'and University, Institute of Physics, P\'azmany P\'eter s\'et\'any 1/A 1117, Budapest, Hungary}
\newcommand{\ELTEastro}{ELTE E\"otv\"os Lor\'and University, Department of Astronomy, P\'azmany P\'eter s\'et\'any 1/A 1117, Budapest, Hungary}
\newcommand{\UMelph}{School of Physics, The University of Melbourne, VIC 3010, Australia}
\newcommand{\astrotd}{ARC Centre of Excellence for All Sky Astrophysics in 3 Dimensions (ASTRO 3D)}
\newcommand{\umdbal}{University of Maryland, Baltimore County, 1000 Hilltop Circle,Baltimore, MD 21250, USA}
\newcommand{\MTAcos}{MTA CSFK Lend\"ulet Near-Field Cosmology Research Group, 1121, Budapest, Konkoly Thege Mikl\'os \'ut 15-17, Hungary}
\newcommand{\coraltower}{Coral Towers Observatory, Cairns, Australia}
\newcommand{\uhawaii}{Institute for Astronomy, University of Hawai'i, 2680 Woodlawn Drive, Honolulu, HI 96822, USA}
\newcommand{\ICECSIC}{Institute of Space Sciences (ICE, CSIC), Campus UAB, Carrer de Can Magrans, s/n, E-08193 Barcelona, Spain.}
\newcommand{\BAERI}{Bay Area Environmental Research Institute, P.O. Box 25, Moffett Field, CA 94035, USA}
\newcommand{\NASAAMES}{NASA Ames Research Center, Moffett Field, CA 94035, USA}
\newcommand{\Rutgers}{Department of Physics and Astronomy, Rutgers the State University of New Jersey, 136 Frelinghuysen Road, Piscataway, NJ 08854, USA}
\newcommand{\Einstein}{NASA Einstein Fellow}
\newcommand{\CIERA}{Center for Interdisciplinary Exploration and Research in Astrophysics (CIERA), Northwestern University, Evanston, IL 60208, USA}
\newcommand{\TelAviv}{The school of Physics and Astronomy, Tel Aviv University, Tel Aviv 69978, Isreal}
\newcommand{\GEMINIobs}{Gemini Observatory, NSF's NOIRLab, Casilla 603, La Serena, Chile}
\newcommand{\Valencia}{Departamento de Astronom\'{\i}a y Astrof\'{\i}sica, Universidad de Valencia, E-46100 Burjassot, Valencia, Spain}
\newcommand{\ValenciaObser}{Observatorio Astron\'omico, Universidad de Valencia, E-46980 Paterna, Valencia, Spain}
\newcommand{\TWNCU}{Graduate Institute of Astronomy, National Central University, 300 Jhongda Road, 32001 Jhongli, Taiwan}
\newcommand{\Carnegie}{The Observatories of the Carnegie Institution for Science, 813 Santa Barbara St., Pasadena, CA 91101, USA}
\newcommand{\umd}{Astronomy Department, University of Maryland, College Park, MD 20742, USA}
\newcommand{\anuobs}{Mt Stromlo Observatory, The Research School of Astronomy and Astrophysics, Australian National University, ACT 2601, Australia}
\newcommand{\anuncpas}{National Centre for the Public Awareness of Science, Australian National University, Canberra, ACT 2611, Australia}
\newcommand{\UTAustin}{Department of Astronomy, University of Texas at Austin, 2515 Speedway, Stop C1400, Austin, TX 78712-1205, USA}
\newcommand{\USzeged}{Department of Optics \& Quantum Electronics, University of Szeged, D\'om t\'er 9, Szeged, 6720 Hungary}
\newcommand{\THCA}{Physics Department and Tsinghua Center for Astrophysics (THCA), Tsinghua University, Beijing 100084, China}
\newcommand{\BJP}{Beijing Planetarium, Beijing Academy of Science and Technology, Beijing 100044, China}
\newcommand{\lco}{Las Cumbres Observatory, 6740 Cortona Drive, Suite 102, Goleta, CA 93117-5575, USA} 
\newcommand{\ucsb}{Department of Physics, University of California, Santa Barbara, CA 93106-9530, USA}
\newcommand{\qub}{Astrophysics Research Centre, School of Mathematics and Physics    , Queen's University Belfast, Belfast BT7 1NN, UK}
\newcommand{\tam}{Mitchell Institute for Fundamental Physics and Department of Physics and Astronomy, Texas A\&M University, College Station, TX 77843.}
\newcommand{\ubirmingham}{Birmingham Institute for Gravitational Wave Astronomy and School of Physics and Astronomy, University of Birmingham, Birmingham B15 2TT, UK}
\newcommand{\UNazar}{School of Engineering and Digital Sciences, Nazarbayev University, Nur-Sultan, Kazakhstan}
\newcommand{\KBTU}{Kazakh-British Technical University, Almaty, Kazakhstan}
\newcommand{\ULison}{CENTRA, Instituto Superior T\'ecnico, Universidade de Lisboa, Av. Rovisco Pais 1, 1049-001 Lisboa, Portugal.}
\newcommand{\UAarhus}{Department of Physics and Astronomy, Aarhus University, Ny Munkegade 120, DK-8000 Aarhus C, Denmark}
\newcommand{\PMO}{Purple Mountain Observatory, Chinese Academy of Sciences, 10 Yuanhua Road, Nanjing 210033, P.R. China}
\newcommand{\uwarsaw}{Astronomical Observatory, University of Warsaw, Al. Ujazdowskie 4, 00-478 Warszawa, Poland}
\newcommand{\CTIO}{Cerro Tololo Inter-American Observatory, NSF's National Optical-Infrared Astronomy Research Laboratory, Casilla 603, La Serena, Chile}
\newcommand{\Uberlin}{Institute of Physics, Humboldt-Universität zu Berlin, Newtonstr. 15, 12489 Berlin, Germany}
\newcommand{\CHEI}{Israel Excellence Fellowship}
\newcommand{\cfa}{Center for Astrophysics \textbar{} Harvard \& Smithsonian, 60 Garden Street, Cambridge, MA 02138-1516}
\newcommand{\notredame}{University of Notre Dame, Notre Dame, IN 46556, USA}

\shorttitle{SN~2018agk a prototypical smooth power-law rise in Kepler}
\shortauthors{Wang et al.}

\begin{document}

\title{SN~2018\lowercase{agk}: A prototypical Type~I\lowercase{a} Supernova with a smooth power-law rise in \kepler\ (K2)}


\author[0000-0001-5233-6989]{Qinan~Wang}\blfootnote{Corresponding author: Qinan~Wang\\ \href{mailto:qwang75@jhu.edu}{qwang75@jhu.edu}}
\affiliation{\JHU}
\author[0000-0002-4410-5387]{Armin~Rest}
\affiliation{\JHU}
\affiliation{\STScI}
\author[0000-0002-0632-8897]{Yossef~Zenati}
\affiliation{\JHU}
\affiliation{\CHEI}
\author[0000-0003-1724-2885]{Ryan~Ridden-Harper}
\affiliation{\JHU}
\author[0000-0001-9494-179X]{Georgios~Dimitriadis}
\affiliation{\UCSC}
\affiliation{\TCD}
\author[0000-0001-6022-0484]{Gautham~Narayan}
\affiliation{\UIUC}
\affiliation{\UIUCCAS}
\author[0000-0002-5814-4061]{V.~Ashley~Villar}
\affiliation{\PSU}
\affiliation{\ICDS}
\affiliation{\IGC}
\author[0000-0002-0629-8931]{Mark~R.~Magee}
\affiliation{\TCD}
\author[0000-0002-2445-5275]{Ryan~J.~Foley}
\affiliation{\UCSC}
\author[0000-0002-3234-8699]{Edward J.Shaya}
\affiliation{\umd}
\author{Peter Garnavich}
\affiliation{\notredame}
\author[0000-0001-7092-9374]{Lifan Wang}
\affiliation{\tam}
\author{Lei Hu}
\affiliation{\PMO}
\author[0000-0002-8585-4544]{Attila B\'odi}
\affiliation{\konkoly}
\affiliation{\MTAcos}
\author{Patrick Armstrong}
\affiliation{\anuobs}
\author[0000-0002-4449-9152]{Katie~Auchettl}
\affiliation{\UCSC}
\affiliation{\UMelph}
\affiliation{\astrotd}
\author[0000-0001-7139-2724]{Thomas~Barclay}
\affiliation{\umdbal}
\author{Geert Barentsen}
\affiliation{\BAERI}
\author{Zs\'ofia~Bogn\'ar}
\affiliation{\konkoly}
\affiliation{\MTAcos}
\author{Joseph Brimacombe}
\affiliation{\coraltower}
\author{Joanna Bulger}
\affiliation{\uhawaii}
\author[0000-0003-0035-6659]{Jamison~Burke}
\affiliation{\ucsb}
\affiliation{\lco}
\author{Peter Challis}
\affiliation{\cfa}
\author{Kenneth Chambers}
\affiliation{\uhawaii}
\author{David A. Coulter}
\affiliation{\UCSC}
\author[0000-0001-9210-9860]{G\'eza Cs\"ornyei}
\affiliation{\konkoly}
\author{Borb\'ala Cseh}
\affiliation{\konkoly}
\author[0000-0001-8857-9843]{Maxime Deckers}
\affiliation{\TCD}
\author[0000-0003-4206-5649]{Jessie L. Dotson}
\affiliation{\NASAAMES}
\author[0000-0002-1296-6887]{Llu\'is Galbany}
\affiliation{\ICECSIC}
\author{Santiago Gonz\'alez-Gait\'an}
\affiliation{\ULison}
\author[0000-0002-1650-1518]{Mariusz Gromadzki}
\affiliation{\uwarsaw}
\author{Michael Gully-Santiago}
\affiliation{\UTAustin}
\author[0000-0002-9415-5219]{Ott\'o Hanyecz}
\affiliation{\konkoly}
\author[0000-0002-3385-8391]{Christina Hedges}
\affiliation{\BAERI}
\affiliation{\NASAAMES}
\author[0000-0002-1125-9187]{Daichi Hiramatsu}
\affiliation{\ucsb}
\affiliation{\lco}
\author{Griffin Hosseinzadeh}
\affiliation{\cfa}
\author[0000-0003-4253-656X]{D. Andrew~Howell}
\affiliation{\ucsb}
\affiliation{\lco}
\author[0000-0002-2532-2853]{Steve~B.~Howell}
\affiliation{\NASAAMES}
\author[0000-0003-1059-9603]{Mark~E.~Huber}
\affiliation{\uhawaii}
\author[0000-0002-2532-2853]{Saurabh~W.~Jha}
\affiliation{\Rutgers}
\author[0000-0002-6230-0151]{David~O.~Jones}
\altaffiliation{\Einstein}
\affiliation{\UCSC}
\author[0000-0002-8770-6764]{R\'eka K\"onyves-T\'oth}
\affiliation{\konkoly}
\author{Csilla Kalup}
\affiliation{\konkoly}
\author{Charles D. Kilpatrick}
\affiliation{\CIERA}
\author{Levente Kriskovics}
\affiliation{\konkoly}
\author{Wenxiong Li}
\affiliation{\TelAviv}
\author{Thomas B Lowe}
\affiliation{\uhawaii}
\author{Steven Margheim}
\affiliation{\GEMINIobs}
\author{Curtis~McCully}
\affiliation{\ucsb}
\affiliation{\lco}
\author{Ayan Mitra}
\affiliation{\UNazar}
\affiliation{\KBTU}
\author{Jose A. Mu\~noz}
\affiliation{\Valencia}
\affiliation{\ValenciaObser}
\author{Matt Nicholl}
\affiliation{\ubirmingham}
\author{Jakob Nordin}
\affiliation{\Uberlin}
\author[0000-0001-5449-2467]{Andr\'as P\'al}
\affiliation{\konkoly}
\affiliation{\ELTEphysics}
\affiliation{\ELTEastro}
\author{Yen-Chen Pan}
\affiliation{\TWNCU}
\author{Anthony L. Piro}
\affiliation{\Carnegie}
\author{Sofia Rest}
\affiliation{\JHU}
\affiliation{\JHUCS}
\author{Jo\~ao Rino-Silvestre}
\affiliation{\ULison}
\author[0000-0002-7559-315X]{C\'esar~Rojas-Bravo}
\affiliation{\UCSC}
\author[0000-0003-0926-3950]{Kriszti\'an S\'arneczky}
\affiliation{\konkoly}
\author{Matthew R. Siebert}
\affiliation{\UCSC}
\author{Stephen J. Smartt}
\affiliation{\qub}
\author[0000-0001-9535-3199]{Ken Smith}
\affiliation{\qub}
\author{\'Ad\'am S\'odor}
\affiliation{\konkoly}
\affiliation{\MTAcos}
\author[0000-0002-5571-1833]{Maximilian D. Stritzinger}
\affiliation{\UAarhus}
\author[0000-0002-3258-1909]{R\'obert Szab\'o}
\affiliation{\konkoly}
\affiliation{\MTAcos}
\affiliation{\ELTEphysics}
\author{R\'{o}bert Szak\'{a}ts}
\affiliation{\konkoly}
\author[0000-0002-4283-5159]{Brad~E.~Tucker}
\affiliation{\anuobs}
\affiliation{\astrotd}
\affiliation{\anuncpas}
\author[0000-0001-8764-7832]{J\'ozsef Vink\'o}
\affiliation{\konkoly}
\affiliation{\ELTEphysics}
\affiliation{\USzeged}
\affiliation{\UTAustin}
\author{Xiaofeng Wang}
\affiliation{\THCA}
\affiliation{\BJP}
\author{J.~Craig~Wheeler}
\affiliation{\UTAustin}
\author[0000-0002-1229-2499]{David~R.~Young}
\affiliation{\qub}
\author{Alfredo~Zenteno}
\affiliation{\CTIO}
\author{KaiCheng Zhang}
\affiliation{\THCA}
\author[0000-0002-4612-5824]{Gabriella Zsidi}
\affiliation{\konkoly}
\affiliation{\ELTEphysics}



\begin{abstract}

\indent We present the 30-min cadence \kepler /K2 light curve of the Type Ia supernova (SN~Ia) SN~2018agk, covering approximately one week before explosion, the full rise phase and the decline until 40 days after peak. We additionally present ground-based observations in multiple bands within the same time range, including the 1-day cadence DECam observations within the first $\sim$5 days after the first light. The \kepler\ early light curve is fully consistent with a single power-law rise, without evidence of any bump feature. 
We compare SN~2018agk with a sample of other SNe~Ia without early excess flux from the literature. We find that SNe~Ia without excess flux have slowly-evolving early colors in a narrow range ($g-i\approx -0.20\pm0.20$ mag) within the first $\sim 10$ days. On the other hand, among SNe~Ia detected with excess, SN~2017cbv and SN~2018oh tend to be bluer, while iPTF16abc's evolution is similar to normal SNe Ia without excess in $g-i$. We further compare the \kepler\ light curve of SN~2018agk with companion-interaction models, and rule out the existence of a typical non-degenerate companion undergoing Roche-lobe overflow at viewing angles smaller than $45^{\circ}$. 
\keywords{: supernovae: general --- supernovae: Type Ia supernova: individual (SN~2018agk)}
\end{abstract}

\section{Introduction} \label{sec:intro}

Type Ia supernovae (SNe Ia) serve as standardizable candles in modern cosmology, and their use as distance indicators led to the discovery of the accelerating expansion of the universe \citep[e.g.,][]{riess1998observational, perlmutter1999measurements}, \redpen{which can be explained by the widely accepted concept of dark energy, or some alternative theories such as modified gravity (e.g. see discussion in \citealp{Joyce2015} and \citealp{Nojiri2017}).} 

\redpen{It has been proposed that SNe~Ia may come from multiple progenitor channels \citep[e.g.,][]{foley2009sn, Polin+19}. \cite{foley2009sn} further reveals that if the light curves of a SNe~Ia subtypes slightly deviate from Phillips relation, there will be a non-negligible systematic bias. Another easy-to-imagine scenario is that, if the rate of different progenitor channels correlates with environment parameters that are different in low- and high-z galaxies, e.g. metallicity, galaxy age and star formation history, then the fraction of these SNe~Ia subtypes will vary with time.This evolution will introduce systematics into the luminosity-distance relation.
In the next decade with the Roman Space Telescope and the Rubin Observatory coming online, the rapidly growing SNe~Ia samples will largely reduce the statistical uncertainties in cosmological measurement as well as improve the photometric calibration, and therefore systematic biases like evolution that are currently secondary will become an important source of uncertainty in cosmological analyses \citep{scolnic2014,scolnic2018complete, riess2019large}.}

In general, SNe~Ia are believed to come from thermonuclear explosions of CO white dwarf (WD) in binaries,
although no ``Branch-normal'' SN~Ia progenitor system has yet been detected in pre-explosion images \citep{maoz2014observational}. The trigger for the thermonuclear runaway (TNR) process in SNe~Ia is thought to occur when the WD reaches or approaches a critical mass limit, known as the Chandrasekhar limit, $M_{ch} \sim 1.4M_\odot$ (in some models TNR could happen in sub-Chandrasekhar scenario, see discussion below). Such process can occur in either the single-degenerate (SD) model \citep[][]{1973ApJ...186.1007W} or double-degenerate (DD) model \citep[][]{IbenTutukov84} or other rare channels \citep[][]{Kushnir+13,KashiSoker11,Pakmor+13,Ruiter20,Pakmor+21}. 
In the common SD model, a central WD accretes matter from a companion star up to $M_{ch}$, and once the carbon ignition occurs near the center, deflagration soon follows and transitions to a detonation. This process produces a SN Ia-like event \citep{Nomoto82a,LivneArnett95,li1997evolution,Kromer+10,Moll+13,maoz2014observational,Soker19,Jha+19}. In the DD model, the TNR process occurs when two CO WDs or CO-HeCO merge via angular momentum loss by radiating gravitational waves \citep[i.e.][]{MariusDan+11,PeretsZenati+19}.

The TNR process can occur at different stages depending on SN~Ia progenitors:  during mass transfer, during the merger, or in the remnant phase after the merger 
\citep{IbenTutukov84,LivneGlanser91,Guillochon+10,Pakmor+13,Fink+10,DanM+14,Kashyap+15,PeretsZenati+19,Polin+19}. The ambiguity in the timing of the TNR process introduces more variation in modelling. For example, \cite{1982ApJ...257..780N} and \cite{1986ApJ...301..601W} proposed the sub-Chandrasekhar Double Detonation (DDet) model, in which a detonation of the helium shell of the WD triggers a denotation of the carbon core. 
Other proposed models include the collision of double WDs \citep{
Kushnir+13}, the explosion of a massive hybrid HeCO WD as a donor in the DD system \citep[][]{zenati+2019,Pakmor+21} and the core-degenerate model \citep[merger of WD and asymptotic giant branch (AGB) star, see ][]{KashiSoker11}. All of these scenarios could give rise to the bulk of SNe~Ia. 

Clues to the progenitor systems of SNe~Ia can be found through early observations of SN~Ia light curves within days after explosion. The canonical `expanding fireball' model predicts that the early rise of SNe~Ia light curve follows $L\propto t^2$ under the assumption that the photospheric temperature roughly remains constant during this period \citep{arnett1982type, riess1998observational}. Previous observations have revealed that the rise of SNe~Ia statistically follow a power-law $L\propto t^\alpha$ with index $\alpha\sim2$, though the index of some SNe~Ia may significantly deviate from $2$ \citep[][]{riess1998observational, olling2015,hayden2010rise,2020ApJ...902...47M}. \redpen{Recently, some models further predict the existence of early excess on top of the power-law rise under some conditions. \cite{kasen2009seeing} 
reveals that in the SD model the interaction between SN ejecta and a non-degenerate companion can produce a blue excess in SN~Ia light curves within days from explosion. In such SD models, H or He features are also expected to be found in late phase spectra \citep{Maeda2014, Botyanszki2018}.}
\redpen{Multiple mechanisms have also been proposed to be able to create a variety of flux excess in early SN~Ia light curves in DD models afterwards.} For example, \cite{2016ApJ...826...96P} shows that shock interaction with circumstellar medium (CSM) can create similar excess to the SD model; 
in the sub-$M_{ch}$ DDet scenario, excess flux can be caused from an initial He-detonation which ejects non-negligible amount of radioactive isotopes on the surface and thus have a relatively red early excess \citep{Kromer+10,Pakmor+13,Kushnir+13,Tanikawa+15,PeretsZenati+19,Ruiter20,Noebauer+17,2020A&A...642A.189M};
due to Fe line blanketing \citep{Maeda+18}.
For the DD scenario, \cite{Levanon+15} and \cite{Levanon+17}
show that if the merger process forms an accretion disk that blows a wind shortly before the explosion, then the collision of the ejecta with this disk-originated matter (DOM) results in an early light excess.
\redpen{Overall, these various models give different predictions on the shape and color of the early flux excess.} These early effects of different models highlight the importance of early observations, as has been proven through combinations of different observational evidence, including light curve shape, color evolution, and early spectra \citep{Jha+19}. 

Continuous high-cadence surveys are a key window to precisely constrain explosion time and search for those potential photometric signatures of progenitors at the very early phase. With its wide field-of-view (FOV) and 30-minute cadence, the \textit{Kepler Space Telescope} \citep[\textit{Kepler};][]{haas2010kepler} was a superb instrument for monitoring thousands of galaxies to observe SNe within hours of their explosion times. During the \textit{Kepler} prime mission, \cite{olling2015} discovered 
3 photometrically classified SNe Ia with coverage from days before explosion to the post-peak phase, which do not have any signatures of early excess. 
The successor of the \Kepler\ prime mission, K2 \citep{2014PASP..126..398H}, had two campaigns (C16~\&~C17) dedicated to the K2 Supernova Cosmology Experiment (K2 SCE), during which the telescope monitored groups of low-redshift galaxies with concurrent ground based observations. This program successfully monitored numerous SNe as well as other extragalactic transients \citep[e.g.][]{rest2018fast}. 

\begin{figure*}[t!]
\begin{center}
\includegraphics[width = \textwidth]{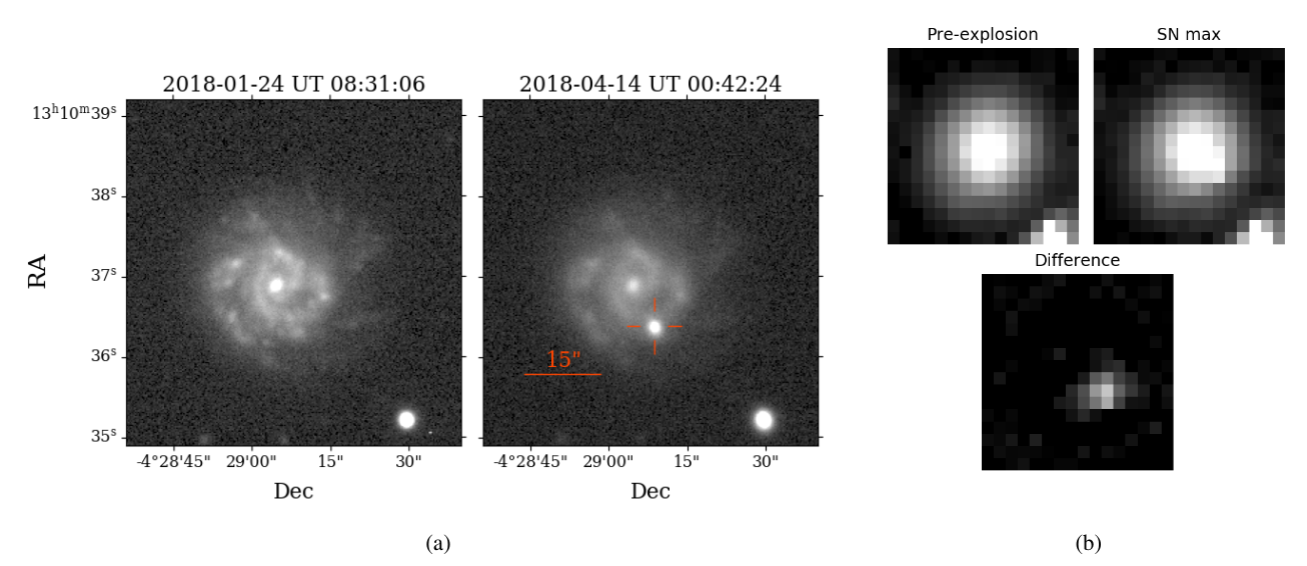}
\caption{\textbf{(a)} The DECam $g$-band image of SN~2018agk and its host galaxy IC 0855 in the pre-explosion stage (left) and $\sim18$ days after B-band maximum (right). The size of two images is $64\arcsec\times64\arcsec$. \textbf{(b)} \kepler\ images of SN~2018agk and its host pre-explosion (top left), peak (top right), and the difference image (bottom). The size of \kepler\ images are $16\times16$ pixels ($64\arcsec\times64\arcsec$).
}
\label{fig:discovery}
\end{center}
\end{figure*}

K2 observed 8 spectroscopically classified SNe~Ia from pre-explosion to post-peak stage (Villar et al in prep). The most remarkable discovery was SN~2018oh which featured a prominent early excess in the \kepler\ light curve for the first $\sim 5$ days after the explosion \citep{dimitriadis2018k2, 2019ApJ...870...12L, shappee2018seeing}. Both SD collision models and DD models with shallow concentration of $^{56}$Ni can approximately reproduce the early \kepler\ light curve shape of SN~2018oh. Alongside early \textit{Kepler}\ photometry, early measurements were also obtained with DECam in $i$. These data suggested an early blue excess, favoring the SD model.
However, the color agreement is contradicted by late-time spectra which showed no evidence of hydrogen or narrow helium emission features that are predicted by the SD model \citep{nebularspec18oh, 2019ApJ...872L..22T}. On the other hand, \cite{2020A&A...642A.189M} also reveals the inconsistency in late-time color and spectra between the observation and prediction of shallow $^{56}$Ni model. Thus, no model can simultaneously match photometric and spectroscopic features of SN~2018oh at this stage. 

Still, our ability to understand the excess flux in SN~Ia such as SN~2018oh is limited from not having a clear template and model to define ``normal'' flux and color evolution -- i.e., a SN~Ia without excess flux. A template SN~Ia with early observations that contains no excess feature is crucial for understanding SNe~Ia with excess flux and distinguishing between different progenitor models. Another problem in analysis is the existence of `dark phase', referring to the time difference between the explosion and the occurrence of observable first light \citep[][]{Piro2013,2020A&A...634A..37M}, and still need to be carefully evaluated in model simulations.

In this paper, we present observations of SN~2018agk, a ``normal'' SN~Ia in both spectroscopic and photometric sense, whose host galaxy was monitored by the K2 SCE from pre-explosion to post-peak stage. We show that the \kepler\ light curve has no signature of early excess features. In addition, SN~2018agk was extensively observed by ground-based observatories in multiple bands, and in particular, DECam observed it at a $\sim 1$~day cadence from pre-explosion to $\sim 4$ days after the first light in $g$ and $i$ bands. In this paper, we focus on the \kepler\ and DECam light curve and color evolution within the first week after first light. With this unique multi-band data, we are able to reveal the intrinsic color evolution of SNe~Ia without excess. Furthermore, we  use the \kepler\ light curve as a baseline to test previous model fitting methods. 

Throughout this paper, observed times are reported in Modified Julian Days (MJDs) while phases, unless where noted, are reported in rest-frame. We adopt the AB magnitude system, unless where noted, and a flat $\Lambda$CDM cosmological model with $H_0 = 73$ km s$^{-1}$ Mpc$^{-1}$ \citep[][]{riess2016, riess2018}.






\section{Observations}

SN~2018agk was discovered with the CTIO 4~m Blanco DECam camera \citep{decam2008, 2015AJ....150..150F} on 2018 Mar 10 at 07:06:06 UTC (MJD 58188.296) with apparent $g$-band magnitude of 20.235 mag \citep{2018TNSTR.346....1R}. The images were taken by the Kepler ExtraGalactic Survey (KEGS) team as part of the ground-based monitoring of the K2 Campaigns~16\&17 \citep{Dotson18}. SN~2018agk occurred at coordinates $\alpha=\rm 13^h10^m36.37^s$, $\delta= -04{\degr}29{\arcmin}08.67{\arcsec}$ (J2000.0) and was spectroscopically classified as a SN~Ia by \citet{2018TNSCR.510....1B}. SN~2018agk is located at a distance of 8.8\arcsec from the center of its host galaxy, IC~0855, which is a spiral galaxy at a redshift of $z = 0.026128$ (see Fig.~\ref{fig:discovery}). 
Throughout this paper, we use the Milky Way extinction of $E(B-V)_{MW}=0.034$ from the extinction map as described in \cite{2011ApJ...737..103S}.

\subsection{Ground-based Photometry}

We obtained ground-based photometry with the CTIO 4-m Blanco telescope with DECam in $g$ and $i$ bands, the Swope 1.0-m telescope at Las Campanas Observatory in $uBVgri$ bands, the 60/90cm Schmidt-telescope on Piszk\'estet{\H o} Mountain Station of Konkoly Observatory in $BVRI$ bands,  the 1.3m telescope at the Cerro Tololo Inter-American Observatory (CTIO) in $RIJHK$ bands, and
the PS1 1.8~m telescope at Haleakala on Maui, Hawai'i in $gri$ bands \citep{2016arXiv161205560C}. We also include photometric data from the Global Supernova Project taken with Las Cumbres Observatory previously published in \cite{2021PASP..133d4002B}.
We downloaded the data products using the NSF NOIRLab DECam Community Pipeline \citep[][]{despipeline2014}. The standard reduction of the Swope images is described in \citet{2018MNRAS.473.4805K}. We reduced the PS1 images using the standard PS1 Image Processing Pipeline (IPP) \citep{2020ApJS..251....3M,2020ApJS..251....5M,2020ApJS..251....6M,2020ApJS..251....4W} which includes standard reductions, astrometric solution, stacking of nightly images, source detection, and photometry. The photometry of all images were calibrated using standard sources from the Pan-STARRS DR1 catalog \citep{2020ApJS..251....7F} in the same field as SN~2018agk and transformed following the Supercal method \citep{2015ApJ...815..117S}. After standard reductions, the {\tt photpipe} pipeline \citep{2005ApJ...634.1103R, 2014ApJ...795...44R} performs difference imaging and transient identification. DECam observed SN~2018agk during the crucial first 5 days after explosion, making it the first SN~Ia that has both a high cadence \Kepler\ light curve and ground-based multi-band observations in its earliest phase. DECam $g$-band images from well before explosion and $\sim 20$ days after peak are shown in Figure~\ref{fig:discovery}(a), and the full multi-band light curves are plotted in Figure~\ref{fig:all_phot}.

\subsection{Kepler Observation}\label{k2observation}

\begin{figure*}[t!]
\begin{center}
\scalebox{1.}
{\includegraphics[width=\textwidth]{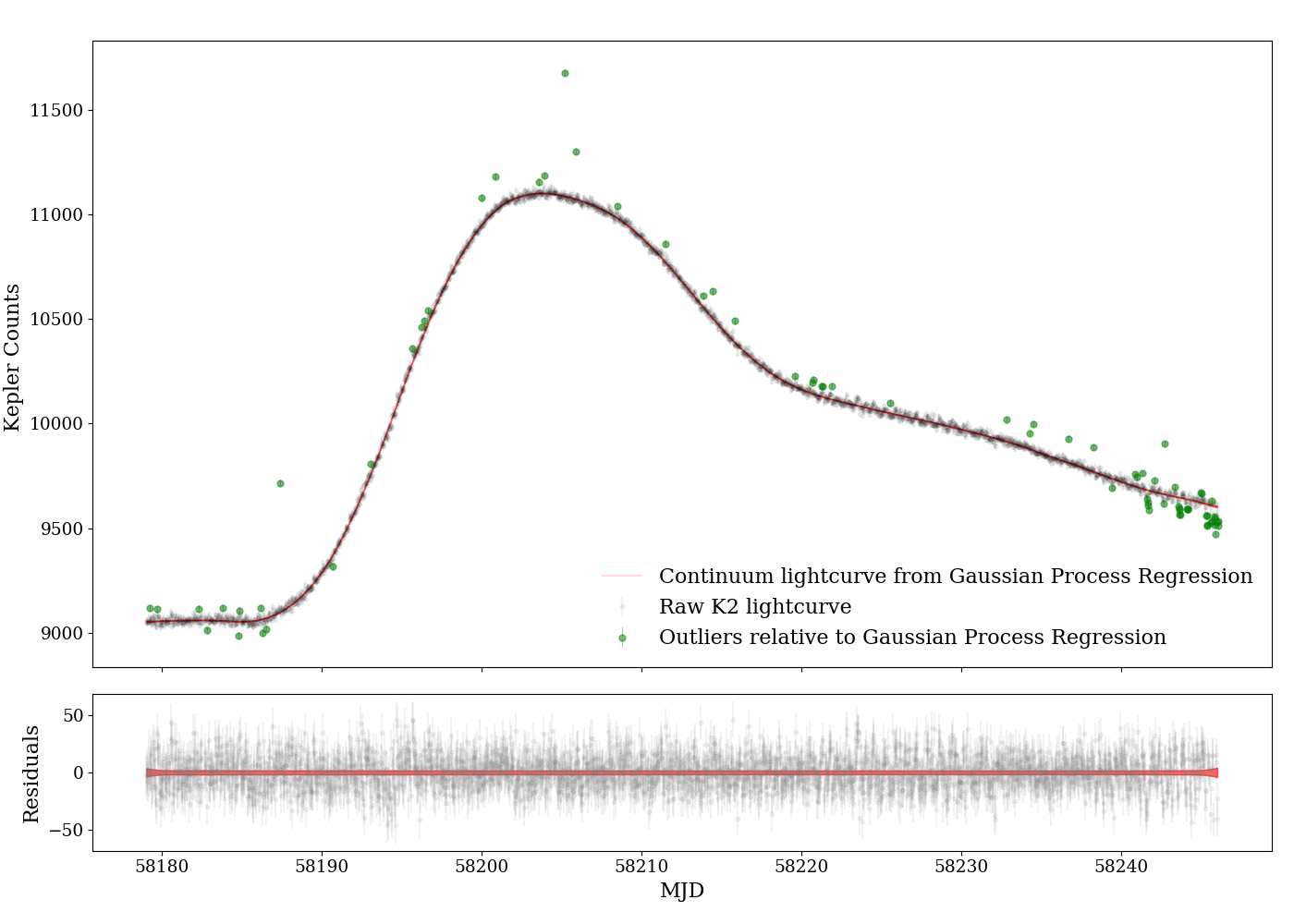}}
\caption{The processed \Kepler\ light curve of SN~2018agk before and after Gaussian Process Regression (GPR) cut (top panel) and residual differences (bottom panel). 
After pipeline reduction, instrumental systematics still have non-negligible influence on \Kepler\ data, e.g. the discontinuities and outliers. We used GPR to remove the outliers and create a smoothed light curve to estimate the peak counts and time for later analysis. We tune the GPR parameter to balance between removing more outliers and keeping data in the fast-evolving early rise phase, and we find the optimal cutoff limit for residuals to be $3\sigma$.}
\label{fig:k2lc}
\end{center}
\end{figure*}

IC~0855 was included as a Campaign 17 target through ``The K2 ExtraGalactic Survey (KEGS) for Transients'' (PI Rest) and the ``Multi-Observatory Monitoring of K2 Supernovae'' (PI Foley) programs as part of the K2 SCE (internal Kepler EPIC ID 228682548). We retrieved the IC~0855 Kepler data through the Mikulski Archive for Space Telescopes (MAST) after the end of K2 Campaign~17.

The unstable pointing of \Kepler\ throughout K2 required special treatment to reduce the data. The \Kepler/K2 mission is characterized by a unique observing strategy that uses spacecraft geometry in conjunction with periodic thruster resets to maintain pointing. This strategy induced a short scale periodic 6 hour `sawtooth pattern', alongside long-term sensitivity trends, due to differential heating  on the spacecraft body and zodiacal background throughout a campaign.

To correct for short and long term trends, we first reduce the data using our K2 reduction pipeline which is described in \cite{2015AJ....150..188S} in more details. This method removes the sawtooth pattern by fitting a third order polynomial in both spatial dimensions of the centroid of the image  to the pattern in the light curve of a fixed 5 x 5 pixel aperture. Long term trends are removed making use of the vectors from our principal-component analysis (PCA) that characterize common simultaneous trends seen in the light curves of all the (assumed) non-varying galaxies observed on the same channel (chip). There were 293 galaxies on the same channel as SN2018agk.
 For supernovae, the applicable coefficients of the PCA vectors and the sawtooth function are estimated using only the times of no supernova flux at the beginning and/or end of the light curve. In this case, we do not have an end anchor point, so the reduced light curve may become unreliable for times significantly past the peak. This is not a significant hindrance to our analysis, as we only use the K2 light curve during the rise. After the long term trends are divided out, we subtract the sawtooth pattern of the galaxy throughout the light curve.  The remainder is the supernova light curve which will generally have a sawtooth pattern of lower amplitude because it is a point source and thus has less of its light shifting in and out of the aperture than the galaxy.  The sawtooth pattern can be scaled down to fit this and removed, but, for this supernova, this was not needed.
 
 
Finally, 
to remove the few outliers, we further model the smooth light curve with a Gaussian Process Regression (GPR) through the \texttt{celerite} package \citep[][]{celerite} and apply a $3-\sigma$ cut, as shown in Fig. \ref{fig:k2lc}. We used a  Mat\'ern 32 kernel with length scales of a few days, which will be insensitive to the short-timescale excursions caused by the K2 thruster resets. In the modelling we explored parameters of the GPR in a reasonable range to balance between rejecting obvious outliers and keeping reliable data in the rise phase for further analysis. 
From the smoothed light curve we find the peak in K2 band occurs at MJD$^{K2}_{max} = 58203.78\pm0.02$.
To estimate noise, we compute the root-mean-squared variation of the background flux before the explosion and then scale it by the square root of the galaxy flux plus the SN flux in the aperture.

\subsection{Ground-based Spectroscopy}

\begin{figure}[t!]
\begin{center}
\hspace*{-0.3in}
\scalebox{1.}
{\includegraphics[width=0.52\textwidth]{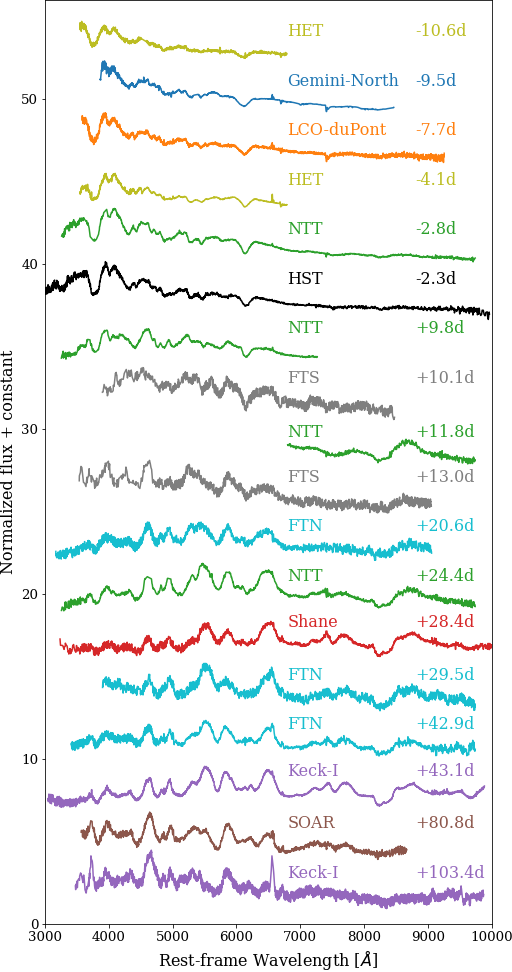}
}
\caption{Spectral series of SN 2018agk. The phases are labeled above each spectra, and spectra from difference instrument are plotted with different colors. The HST spectra taken with three gratings on around MJD 58201.8 ($\sim 2.3$ days before B-band maximum) have been combined into one. All the spectra have been shifted for display purpose.}
\label{fig:spec}
\end{center}
\end{figure}
We obtained a total of 17 spectra for SN~2018agk from ground-based observatories: one spectrum with the Gemini Multi-Object Spectrograph \citep[GMOS;][]{GMOS} on the Gemini-North telescope (GN-2018A-LP-13, PI: Garnavich); 4 spectra with ESO Faint Object Spectrograph and Camera \citep[EFOSC2;][]{EFOSC2} on the ESO New Technology Telescope (as part of the ePESSTO survey, \citealt{Smartt15AA}); one spectrum with the Kast spectrograph \citep[KAST;][]{KAST} on the Lick Shane telescope (2018A-S023, PI: Foley); 2 spectra with the Low-Resolution Imaging Spectrometer \citep[LRIS;][]{LRIS} on the Keck I telescope (U240, PI: Foley); one spectrum with the Goodman High Throughput Spectrograph \citep{Clemens04SPIE} at the Southern Astrophysical Research Telescope (2018A-0277, PI: Foley); 5 optical spectra with the twin FLOYDS spectrographs mounted on Las Cumbres Observatory's 2-m Faulkes Telescopes North (FTN) at Haleakala Observatory in Hawai'i and Faulkes Telescopes South (FTS) at Siding Spring Observatory in Australia; and 2 spectra with the Low Resolution Spectrograph-2 (LRS2) \citep{Chonis2016} on the 10m Hobby-Eberly Telescope (HET) at McDonald Observatory, where the UV and Orange arms of the Blue Integral Field Unit were applied resulting in a spectrum between 3640 - 6970 \AA. We additionally use the publicly available classification spectrum, posted on the Transient Name Server (TNS), obtained with the Wide-Field CCD (WFCCD) of the 2.5m du Pont Telescope at Las Campanas Observatory \citep[][]{2018TNSCR.510....1B}.

The spectra were reduced using standard IRAF/PYRAF and python routines for bias/overscan subtractions and flat fielding. The wavelength solution was derived using arc lamps while the final flux calibration and telluric lines removal were performed using spectro-photometric standard star spectra.

Spectra from Gemini-North, Shane KAST, Keck I LRIS, and SOAR were reduced using standard \textsc{IRAF/PYRAF}\footnote{IRAF was distributed by the National Optical Astronomy Observatory, which was managed by the Association of Universities for Research in Astronomy (AURA) under a cooperative agreement with the National Science Foundation} and python routines for bias/overscan subtractions and flat fielding. The wavelength solution was derived using arc lamps while the final flux calibration and telluric lines removal were performed using spectro-photometric standard star spectra. 
The EFOSC2 spectra were reduced in a similar manner, with the aid of the PESSTO pipeline\footnote{\url{https://github.com/svalenti/pessto}}. The HET LRS2 spectra were reduced through custom-developed IRAF scripts, see \citet{2020ApJ...902...46Y} for more details on the LRS2 IFU spectrograph and the reduction process. The FLOYDS spectra from Las Cumbres Observatory's FTS and FTN were reduced using standard IRAF tasks as described in \citet{Valenti2014}. Table ~\ref{tab:spec} in Appendix summarizes our spectroscopic observations and Figure~\ref{fig:spec} shows the evolution of the spectra.

\subsection{Hubble Space Telescope Observations}
We triggered Hubble Space Telescope ($HST$) followup of SN2018agk on 2018 March 16 (GO-15274, PI Garnavich) and the first exposure began March 19 17:09 (UT). However, due to a $HST$ gyro glitch, the visit failed after the first exposure. A full complement of observations were obtained on the second visit five days later. During both visits, Space Telescope Imaging Spectrograph (STIS) spectra were obtained using the G230L grating and 0.2 arcsec wide slit using the NUV-MAMA detector. On the second visit, CCD spectra using the G430L and G750L gratings were obtained providing wavelength coverage from 170~nm to 1010~nm. The spectra were extracted, wavelength corrected and flux calibrated through the standard STIS pipeline, and are plotted in Fig.~\ref{fig:uv} in contrast to the HST UV spectra of SN~2011fe taken around the similar phase.
\begin{figure}[t!]
\begin{center}
\scalebox{1.}
{\includegraphics[width=0.5\textwidth]{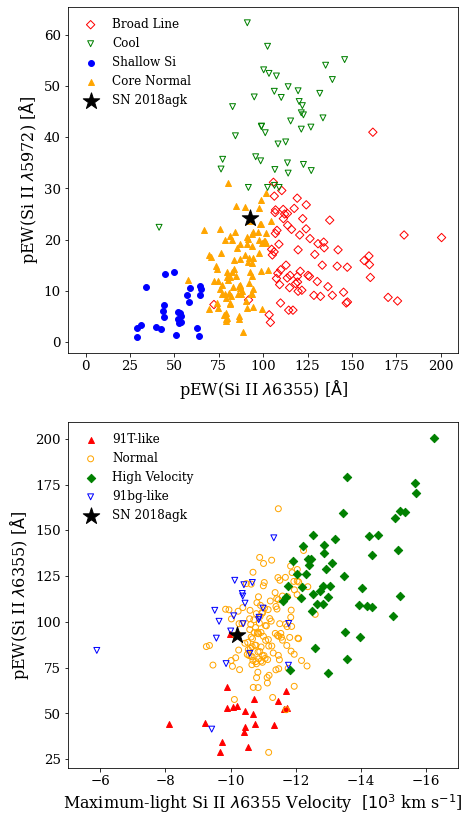}}
\caption{Classification diagram of the subtypes of SNe Ia as defined by \cite{branch2006} and \cite{wang2009}. The pEW and velocity of Si lines in the HST spectra of SN 2018agk at 2.3 days before B-band maximum (black star) indicates it is a normal SN Ia. }
\label{fig:spec_config}
\end{center}
\end{figure}

\begin{figure*}[t!]
\begin{center}
\hspace*{-0.3in}
\scalebox{1.}
{\includegraphics[width=0.9\textwidth]{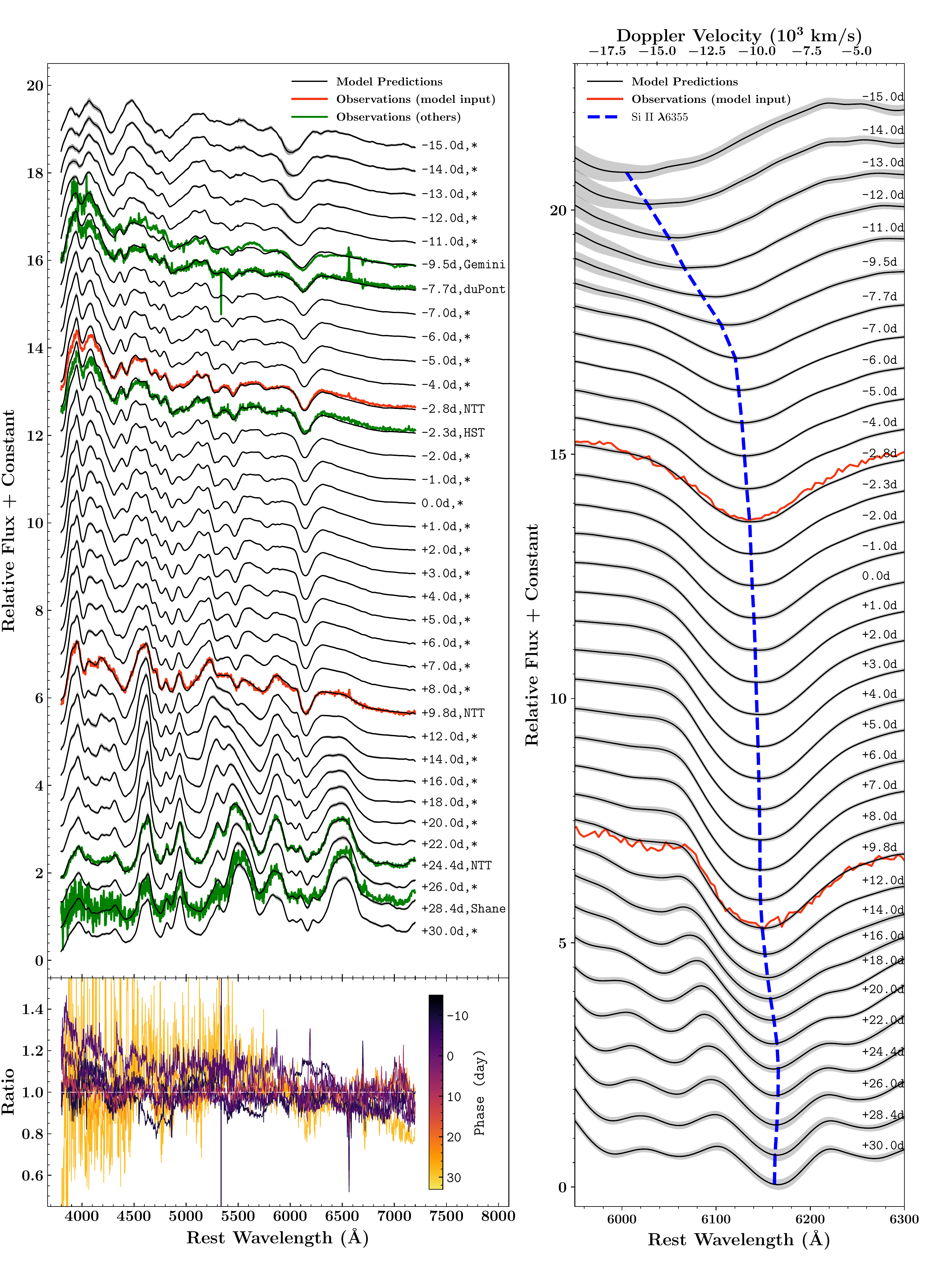}
}
\caption{\textbf{Left:} The spectral sequence of SN~2018agk projected by the model based on Long Short-Term Memory (LSTM) neural networks \citep[][]{lstmnn}. The two NTT spectra around the peak with highest S/N (red) are used to build the projections (black), and the other high S/N spectra (green) are included for fidelity test. The grey regions indicate the $2-\sigma$ standard deviation of the projection. All of the spectra have been shifted in vertical direction, and labeled with corresponding phase and instrument. Bottom panel shows the ratios of the observations to the projections. The precision of the projections is around a few percent in general. \textbf{Right:} The evolution of Si II $\lambda6355$ of LSTM neural network projections of SN~2018agk. Blue dashed line indicates the velocity evolution trend. For clarity, only two spectra used for building the projections are overplotted in red.
}
\label{fig:v_si1}
\end{center}
\end{figure*}

\begin{figure}[t!]
\begin{center}
\hspace*{-0.3in}
\scalebox{1.}
{\includegraphics[width=0.55\textwidth]{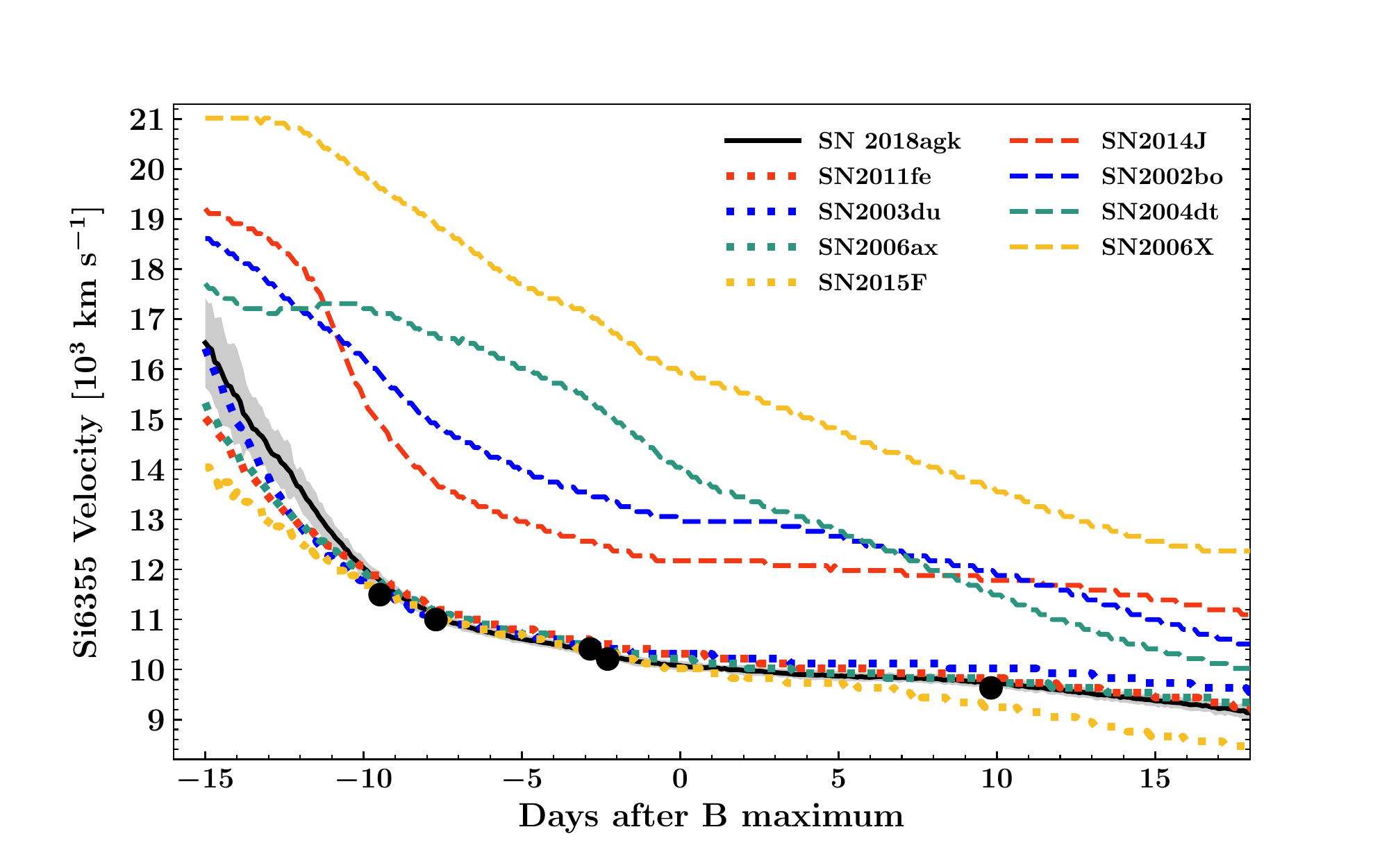}
}
\caption{Evolution of Si II $\lambda$6355 velocity of SN~2018agk. The black dots denote the direct measurements from the observations, and the black line and grey region denote the projections and uncertainty range from the model based on LSTM neural network presented in \cite{lstmnn}. The evolution of Si II velocity of normal SNe~Ia (colored dotted lines) and high-velocity SNe~Ia (colored dashed line) projected from the same LSTM neural network have also been plotted for comparison.
}
\label{fig:v_si2}
\end{center}
\end{figure}

\section{Analysis}

\subsection{Spectroscopy}\label{sec:spec_analysis}

\cite{branch2006} and \cite{wang2009} showed that the pseudo-equivalent width (pEW) of the Si absorption lines (Si II $\lambda$6355 and Si II $\lambda$5972) and the velocity of Si II $\lambda$6355 line around peak can be used to classify SNe Ia into different branches. Here we applied this classification schema to SN 2018agk. To estimate pEW, we first define the pseudo-continuum $f_c(\lambda)$ by the linear curve connecting local maxima at the edge of the absorption features that doesn't intersect the spectral features. Then we integrate the flux normalized by the pseudo-continuum through the formula 

\begin{equation}
    \text{pEW} = \int \left( \frac{f_c(\lambda_i)-f(\lambda_i)}{f_c(\lambda_i)}\right) dx
\end{equation}
where $f(\lambda_i)$ is the measured flux. We perform the integration using \texttt{scipy.integrate.quad}. We also measure the velocity of Si II 6355 {\AA} line by the blueshift of absorption minimum in the spectrum normalized by pseudo-continuum.

We do the calculation on the HST spectrum taken on MJD58201.90, $\sim 2.2$ days prior to the B-band maximum, and measure the pEW of Si II $\lambda$6355 to be $92.621\pm 0.009${\AA} and  pEW of Si II $\lambda$5972 to be $24.1 \pm 0.1${\AA}. The velocity of Si II $\lambda$6355 is $-10200\pm100$km/s. The Branch diagram is plotted in Fig.~\ref{fig:spec_config}, in comparison to the sample from \cite{2012AJ....143..126B}. It clearly shows that SN~2018agk matches the features of branch normal SNe~Ia and does not have any peculiar features.

\begin{figure}[t!]
\begin{center}
\scalebox{1.}
{\includegraphics[width=0.48\textwidth]{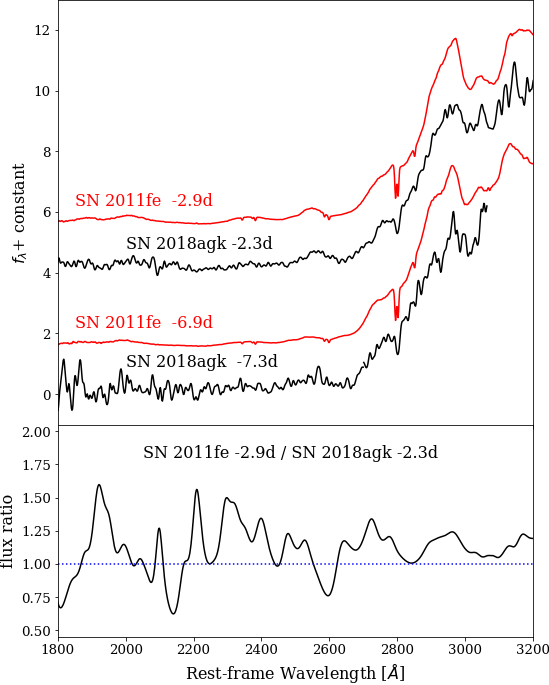}
}
\caption{\textbf{Top:} {\textit{HST}} UV spectra of SNe~2011fe (red) and 2018agk (black)  at similar phases of around $-2.5$ (top) and $-7$~days (bottom). The spectra have been dereddened and normalized to their flux around $3000$\AA\ and also smoothed by a Gaussian filter with $\sigma = 1.5$\AA\ for presentation purpose. \textbf{Bottom:} Flux ratio of the $\sim -2.5$-day SNe~2011fe and 2018agk UV spectra normalized to their flux around $4000$\AA . Due to the relatively low S/N in UV region, the ratio has been smoothed by a Gaussian filter with $\sigma = 12$\AA\ for presentation purpose. }
\label{fig:uv}
\end{center}
\end{figure}

\begin{figure*}[t!]
    \centering
    \includegraphics[width=\linewidth]{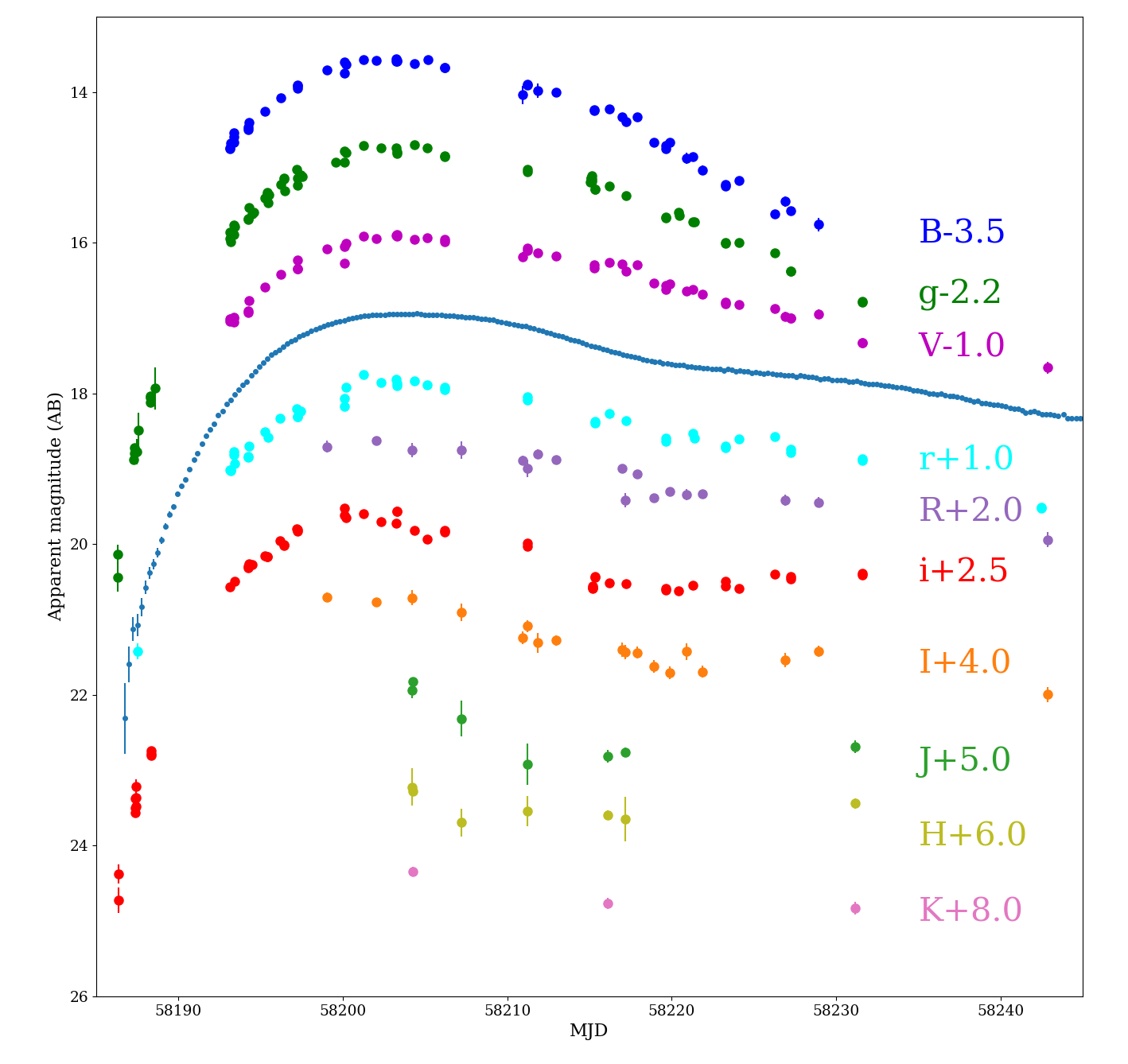}
    \caption{The optical and near-infrared light curves of SN~2018agk, including the 6-hour binned \kepler\ light curve. The light curves are shifted vertically for clarity of display.
    }
    \label{fig:all_phot}
\end{figure*}

The spectroscopic evolution of SNe~Ia forms a rather homogeneous family which allows any individual SN~Ia to be studied based on knowledge of the entire SN~Ia sample. The data-driven model of the SNe~Ia developed by \cite{lstmnn} is applied to the spectroscopic data of SN~2018agk. This model is based on the Long Short-Term Memory (LSTM) neural networks. The model can construct a complete spectral sequence of an SN~Ia using spectroscopic data around optical maximum. For normal SNe~Ia, the flux level of the entire spectral sequence can be reconstructed to a precision of better than 7\%. The evolution of spectral features can also be predicted accurately. The left panel of Figure \ref{fig:v_si1} shows the LSTM neural network projection of the spectral sequence of the SN 2018agk. We picked two spectra with highest S/N around peak, plotted in red color, to build the projections which are plotted as black lines. The rest of spectra with high S/N plotted in green are used to test the fidelity of the neural network projects. The bottom panel shows the ratios of the observations to the neural network projections. The predictions are good to a few percent in general. The phases of the spectra were shown in the color bars on the right. It can be seen that the spectral matches up to the earliest spectroscopic observations (day -9.5 from optical maximum) and down to around 1 month after the optical maximum can be very well reconstructed using only the spectra taken around and after optical maximum (day -2.8 and 9.8 from optical maximum). The success of this reconstruction again supports the notion that SN~2018agk is a member of well-observed SNe~Ia that are used to build the LSTM neural network models. 
We further utilize the LSTM neural network projections to study the evolution of the blueshifts of the Si II $\lambda6355$ features, as shown in the right panel of Figure \ref{fig:v_si1}. The evolution trend of Si II $\lambda6355$ measured from the projected spectra of SN~2018agk has been plotted in Figure \ref{fig:v_si2} against the projections of other SNe~Ia from the same LSTM neural networks. The black line shows the velocity of SN~2018agk measured from the projected spectra and the black solid dots show the measured velocity from the observed spectra. In this Figure we also plot normal SNe~Ia (dotted lines) and high velocity group (dashed lines). SN~2018agk shows strong similarity with normal SNe~Ia group and validates our conclusion that SN~2018agk is a typical normal-branch SN~Ia.

We estimate the host extinction in the line of sight using the method from \citet{Poznanski12MNRAS}. After correcting our highest resolution spectra for redshift, we fit the Na I D ($\lambda5890$) doublet and measure equivalent width $EW=0.75\pm0.15$ {\AA}. Using equation 9 from \citet{Poznanski12MNRAS} we estimate $E(B-V)_{\mathrm{host}}=0.11\pm0.05$.

Differences in SN~Ia progenitor metallicity are theoretically expected to affect both the peak luminosity and UV spectral-energy distribution (SED) of a SN~Ia while having minimal impact on the optical SED \citep[e.g.,][]{hoeflich1999, lentz2000}.
The predicted trends have been observed in SN~Ia data \citep{Foley2013, Foley2016, foley2020, graham2015}, showing a difference in UV continuum that correlates with shape-corrected luminosity \citep{foley2020} and host-galaxy metallicity \citep{Pan2020}.
Since the exact model SED depends on the exact progenitor and explosion model, \citet{Foley2013} used the flux ratio between \citet{lentz2000} models with different metallicities compared to the flux ratio between different SNe to determine the relative metallicity of the SNe.
We compare the reddening-corrected UV spectra of SN~2018agk at two epochs compared to phase-matched spectra of SN~2011fe, a SN~Ia with a similar light-curve shape as SN~2018agk (Fig.~\ref{fig:uv}).  
In general, the UV spectra of two SNe Ia share similar continuum levels and spectral features, with the flux ratio being close to unity at both epochs with variations generally $<$50\%, suggesting good agreement given differences in the line features.
Considering that SNe~2011fe and 2018agk have similar ejecta velocity evolutions as shown in Fig.~\ref{fig:v_si2}, similar $\Delta m_{15}$ and rise time (see Section \ref{photometric} and \ref{section:modellc}) and the similarity in their UV SEDs, we can qualitatively conclude that the progenitors of the two SNe~Ia had similar metallicity.

\begin{figure}[t!]
    \centering
    \includegraphics[width=1.1\linewidth]{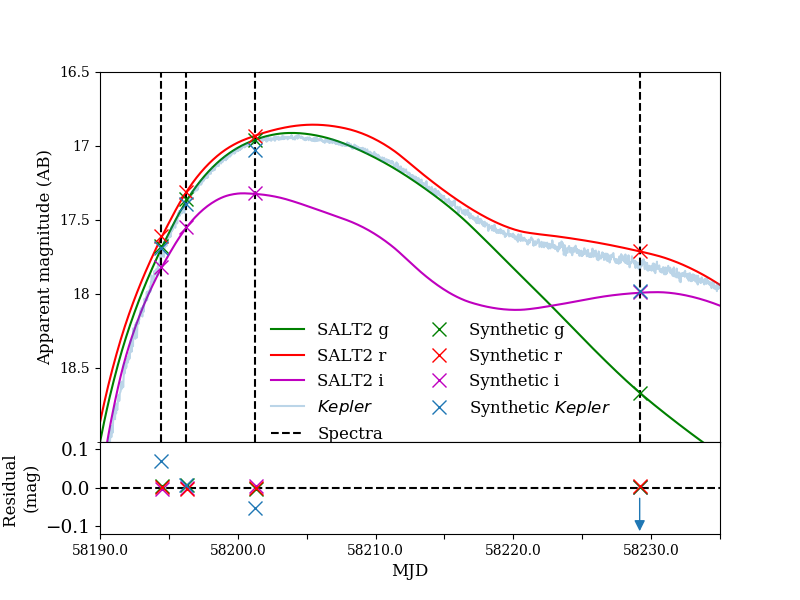}
    \caption{Calibration of \Kepler\ photometric zero-point with synthetic magnitudes and SALT2 fitting results. The black dash lines mark the time of spectra used to calculate the synthetic magnitude, and colored lines are the fitted SALT2 model in $gri$ bands, as described in Section~\ref{photometric}. Spectra used to generate synthetic magnitudes are calibrated to the SALT2 model, from which the \Kepler\ zero-point $zp$ is obtained. The large offset ($\sim0.2$~mag) of \Kepler\ observed and synthetic \Kepler\ magnitudes around $\rm MJD\approx58229.2$ indicates the breakdown of the \Kepler\ data reduction method at late time. Parameters from SALT2 fitting have been listed in Table~\ref{tab:salt2}.}
    \label{fig:salt2zp}
\end{figure}

\subsection{Light Curve Fitting}\label{photometric}

We use the \texttt{sncosmo} package to fit Swope and DECam data with the SALT2 model\footnote{Due to the limited wavelength range, SALT2 can not fit to Kepler light curve of nearby SNe Ia within certain redshift range, including SN~2018agk.}. For the Swope {\textit{B}} and {\textit{u}} band, it is difficult to obtain an accurate photometric calibration using the PS1 $griz$ as other bands, due to the lack of wavelength overlap and poor filter shape match. Therefore we exclude the Swope {\textit{B}} and {\textit{u}} band light curves from the SALT2 fit. When fitting, we fix parameters that are already determined via other methods: redshift, Milky Way and host extinction. The results are listed in Table \ref{tab:salt2}. The time of peak in B-band is MJD$^{B}_{peak} = 58204.178$, $\sim 0.2$ days earlier than the \kepler\ maximum.  The absolute peak magnitude in B-band is measured to be $M_B = -18.950\pm0.002$, and the decline rate in B-band is $\Delta m_{15} = 1.074\pm0.003$. All the photometric data are plotted in Figure \ref{fig:all_phot} and part of the SALT2 light curve are plotted in Figure \ref{fig:salt2zp} . All of SN~2018agk's SALT2 parameters are well within the range of a normal SN~Ia, reinforcing that it is a typical normal SN~Ia \citep{scolnic2018complete}. 

\begin{table}[!]
\centering
\scriptsize
\begin{tabular}{c c c c c} 
\hline
$t^{B}_{peak}$ & $x_0$ & $x_1$ & c \\ [0.3ex] 
\hline\hline
$58204.178\pm 0.008$ & $(4.540\pm0.008)\times10^{-3}$ & $0.220\pm 0.015$ & $0.0168\pm 0.0011$ \\
\hline
\end{tabular}
\caption{SALT2 fitting parameters}
\label{tab:salt2}
\end{table}

\subsection{Kepler Photometric Calibration}\label{calibration}
Kepler features a broadband filter that covers the $g, r$ and $i$ bands with a wavelength range of 4183.66 to 9050.23~{\AA}. We calibrate the observed Kepler counts to physical AB magnitudes with the SN~2018agk spectra that fully cover the Kepler bandpass taken on MJD 58194.45, 58196.26, 58201.26 and 58229.17 (see Figure~\ref{fig:salt2zp}). Since there are few ground based observations that coincide with the spectra, we normalise the spectra by mangling it to the SALT2 model $gri$ magnitudes using the \texttt{mangle\_spectrum2} routine from the \texttt{SNooPy} package \citep{burns2011}.
We calculate the synthetic Kepler magnitude for each spectra using the Kepler bandpass available on Spanish Virtual Observatory (SVO) \citep[][]{2012ivoa.rept.1015R,2020sea..confE.182R} and algorithms in the \texttt{pysynphot} package \citep[][]{2013ascl.soft03023S}. 

With the synthetic \Kepler\ magnitudes, we then calculate the Kepler zero-point using the first three values from rise to peak. We limit the sample to these three points as at late times the \Kepler\ data reduction method breaks down leading to unrealistic count levels (see Figure~\ref{fig:salt2zp}). To avoid biasing from residual saw tooth variability in the Kepler light curve we heavily smooth the light curve with the Savitzky-Golay smoothing method, using a 3rd order polynomial and a window size of 201 frames (100.5~hours). We calculate the zero-point according to:
\begin{eqnarray}
zp=Kp_{syn} + 2.5log(C),
\end{eqnarray}
where $Kp_{syn}$ is the synthetic Kepler magnitude and $C$ is the observed smoothed counts. We find all three points are consistent with $zp=25.219\pm0.045$.



\begin{figure*}[t!]
\begin{center}
\hspace*{-0.1in}
\scalebox{1.}
{\includegraphics[width=0.7\textwidth]{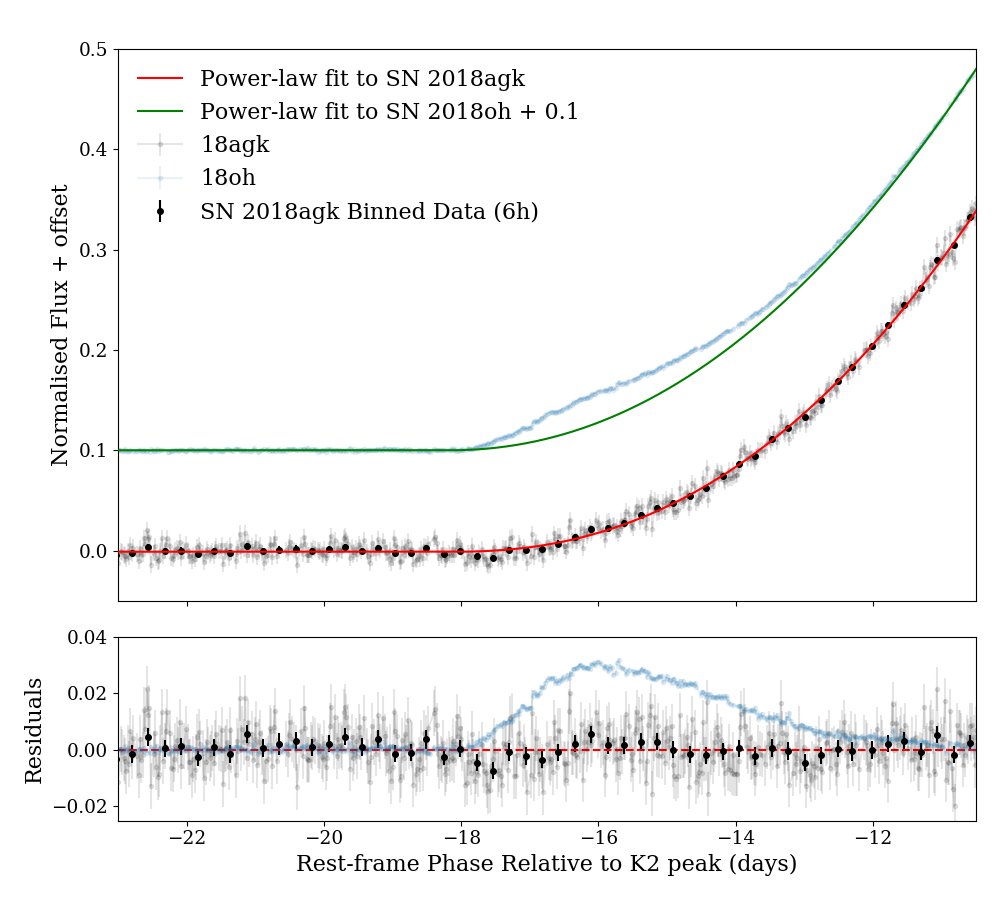}}
\caption{\textbf{Top:} Comparison between the early rise and power law fit for SNe~2018agk and 2018oh. Phases are relative to the \redpen{time of peak brightness} in \kepler\ band, \redpen{and light curves have been normalized by their peak magnitudes}. The raw \kepler\ light curves with 30-minute cadence of SNe~2018agk and 2018oh are plotted as grey and blue points respectively. The 6-hour average for SN~2018agk is shown in black points, and the power-law fits for SNe~2018agk and 2018oh are shown as red and green lines, respectively. For SN~2018oh, we used the power-law parameters from the two-component fit (power-law + skewed Gaussian) described in \cite{dimitriadis2018k2}. 
\textbf{Bottom:} Residuals of the SN~2018agk fit compared with the excess of SN~2018oh relative to its power-law component in the fit. \redpen{SN~2018agk is well fit by the power-law rise without excess flux, whereas the excess is clear for SN~2018oh.} For clarity, the $t^2$ and double power-law fits to SN~2018agk are not included because they are almost indistinguishable with the single power-law fit in most region.
}
\label{fig:RISEcomp}
\end{center}
\end{figure*}

\subsection{Modeling the Rise of the Light Curve}\label{section:modellc}

\begin{table*}[t!]
\centering
 \begin{tabular}{c c c c c c c c c c} 
 \hline
 model & $t_{fl}$ & $t_{fl}^\prime$ & $\alpha$ & $\alpha^\prime$ & $A(\times10^{3})$ & $A^{\prime}(\times10^{3})$ & $f_0(\times10^{4})$ & $\chi^2$ & BIC\\ [0.3ex] 
 \hline\hline

$t^2$ rise & $-17.53\pm0.03$ & - & 2 & - &$6.79\pm0.03$& - & $-3.2\pm4.2$& 497.0 & $-3880.5$\\
$t^\alpha$ rise & $-18.09\pm0.16$ & - & $2.25\pm0.07$ & - &$3.53\pm0.06$& - & $-7.6\pm4.3$ & 478.5 & $-3896.7$\\
double $t^\alpha$rise & $-14.94\pm0.31$ & $-17.05\pm0.18$ & $1.79\pm0.11$ & $1.14\pm0.20$ &$11.4\pm2.8$& $20.3\pm4.6$ & $-7.3\pm4.1$ & $462.6$ & $2257.3$\\

\hline
\end{tabular}
\caption{Different power-law fitting results on \Kepler\ light curve of SN~2018agk. }
\label{tab:fitresult}
\end{table*}

\Kepler\ provided a single broad-band light curve covering the pre-explosion and rising phase of SN~2018agk starting from MJD 58179.05. With these calibrated data, we determine the onset \redpen{of the supernova light curve as follows}: we define the interval from the start of \textit{Kepler} observation to $22$ days before $MJD^{K2}_{max}$ as the quiescent background and calculate $3-\sigma$ clipped weighted average and uncertainty $\sigma$ in this interval, and then we \redpen{mark} the time of K2 detection as the time when the GPR smoothed flux (see Section \ref{k2observation}) rises to $1-\sigma$ above the median flux of the background. We calculate MJD$_{det}^{K2} = 58186.63$
($17.14$ days before $t^{peak}_{Kepler}$)
 as the result. In comparison, the first detection from ground was at MJD$ = 58186.29\pm0.02$ in DECam-g band, $\sim 0.3$ 
days earlier than MJD$_{det}^{K2}$.


 The early rise of our flux-calibrated light curve for SN~2018agk is shown in Fig. \ref{fig:RISEcomp}. We also include the rise of SN~2018oh, which has a clear excess in the first $\sim 5$ days after first light \citep{dimitriadis2018k2,shappee2018seeing,2019ApJ...870...12L}. The comparison shows that there is no large excess in SN~2018agk's \Kepler\ light curve. We then use a series of different power-law models to fit the light curve of SN~2018agk. The form of the basic power-law model is defined as
\begin{equation}
\begin{aligned}
    f(t) & = H(t_{fl})A(t-t_{fl})^\alpha, 
\end{aligned}
\end{equation}
where $A$ \redpen{is} the scale parameter, $t_{fl}$ is the time of the first light, and $H(t_{fl})$ is the Heaviside function such that $H(t_{fl})=0$ for $t<t_{fl}$ and $H(t_{fl}) = 1$ otherwise. We set the baseline flux equal to 0 in the fitting since the background flux has been subtracted out in normalization. We replicate the method used in \citet{olling2015} \& \citet{dimitriadis2018k2} and fit to data in the range of $25$ days before peak until the \Kepler\ flux reaches $35\%$ of the peak flux ($\sim$ MJD$58193.18$ or $\sim 11$ days before peak), using the Python package \texttt{scipy.optimize.least\_squares}.

To test whether SN~2018agk follows traditional `expanding fireball' model or a more general power-law rise, and to examine whether a bump feature exists in SN~2018agk, we fit the \Kepler\ light curve with three models: a fixed single power-law, a $t^2$ rise leaving $t_{fl}$ as a free parameter, and a two power-law with different $t_{fl}$, $t_{fl}^\prime$ and $\alpha$, $\alpha^\prime$. 
In order to judge the goodness of power-law fits with different number of parameters, we apply the Bayesian Information Criteria (BIC), which is a model selection technique based on the goodness-of-fit and number of model parameters. As presented in \citet{priestley1981spectral}, the BIC for a given fit can be calculated as:
\begin{equation}
    \texttt{BIC} = k \ln (n)+n \ln(\hat{\sigma}^2)
\end{equation}
\noindent where $k$ is the number of parameters in the model, $n$ is the number of datapoints in the sample, and $\hat{\sigma}^2 = \chi^2/n$ is the variance, which is defined as the normalized mean of the squared difference between the data and the model. The model with smaller BIC is preferred in terms of balance between the agreement with data and complexity of model. Normally, a difference of $>6$ between BIC values indicates a statistically significant advantage.

The fitting results are listed in Table~\ref{tab:fitresult}. The BIC of single power-law fit is smaller than that of $t^2$ fit by $>8$ and thus is significantly preferrable. Compared with the single power-law fit, double power-law fit does not significantly improve the fitting quality, as $\chi^2$ of two fittings are similar, and because of larger degree of freedom, its BIC is much larger. Thus, we can conclude that the \Kepler\ light curve of SN~2018agk can be well-fitted by a single power-law and there is no evidence of excess-like features at the earliest stage. The inferred time of first light is MJD $58185.22$, $\sim 1$~days before the first detection in DECam and \kepler\ as calculated in Section~\ref{photometric}.

This result can be further validated when we compare the fit of SN~2018agk with that of SN~2018oh, as shown in Fig.~\ref{fig:RISEcomp}. For SN~2018oh we used the result of two component fit (power-law component for supernova flux and skewed Gaussian profile for excess) as described in \cite{dimitriadis2018k2}. Comparing the residuals to the power law fits shown in the lower panel make it clear that the excess of SN~2018oh is well beyond the uncertainty limit of the binned light curve of SN~2018agk.

\begin{figure*}[t!]
\begin{center}
\scalebox{1.}
{\includegraphics[width=\textwidth]{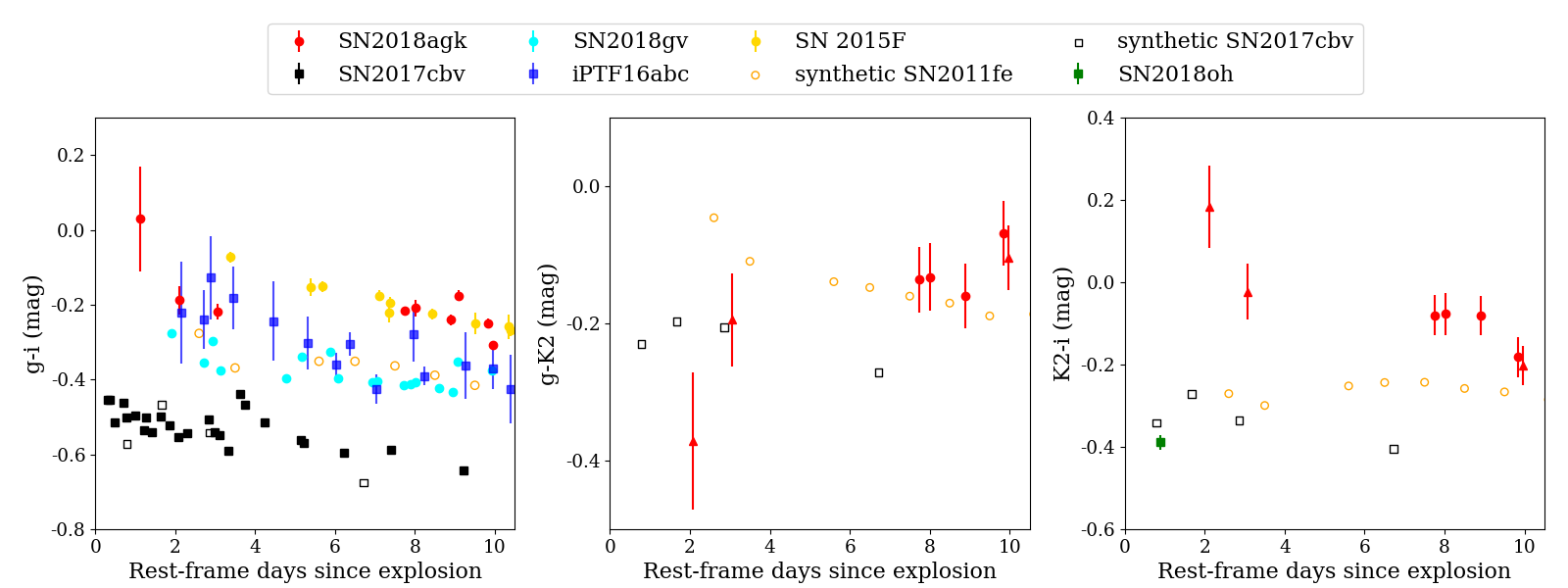}}
\caption{Color evolution of SN~2018agk in $g-i$ (left), $g-Kepler$ (middle) and $Kepler-i$ (right). We also include other SN~Ia with early observations, such as photometric measurements of SN 2015F, SN 2017cbv, SN 2018oh and SN 2018gv (in $g$ and $i$ band), alongside synthetic magnitude of SN2011fe and SN 2017cbv calculated from photometrically calibrated spectra. All photometric measurements include uncertainties, but may be too small to be seen. The SNe~Ia without bump and iPTF16abc concentrate in $g-i$ band within a range of $-0.2\pm0.2$, while SN~2017cbv is bluer than this normal SNe~Ia group. All SNe~Ia in our sample have relatively stable color evolution in the first $\sim 10$ days.}
\label{fig:color evolution} 
\end{center}
\end{figure*}

\subsection{Color Evolution}

Using the previous fits we calculate the range of color evolution for SN~2018agk. The results in the crucial early phase ($\sim 10$ days after $t_{fl}$) are plotted in Fig.~\ref{fig:color evolution}, along with DECam and Swope measurement in both g- and i-band in $\lesssim 4$ days and $\gtrsim 7$ days relative to $t_{fl}$ respectively. 
Overall, the color of SN~2018agk in these 3 bands is relatively constant in the first $\sim 10$ days, with variance $< 0.1$ mag excluding the first observation due to the large uncertainty in $g-i$ band. A linear fit to $g-i$ of SN~2018agk gives a change rate of $0.007\pm0.003$~$\rm mag\;day^{-1}$. Qualitatively speaking, this result agrees with the assumption of the fireball model that the photospheric temperature approximately remains constant when it expands. 

We also compare SN~2018agk with other SNe~Ia that have early photometric measurements or spectra series. Our sample includes SN2011fe \citep{2011fe-specphot}, SN~2015F \citep[][]{SN2015F}, iPTF16abc \citep[][]{iptf16abc}, SN~2017cbv \citep{hosseinzadeh2017early}, SN~2018gv \citep{2020ApJ...902...47M}, SN~2018oh \citep{dimitriadis2018k2, 2019ApJ...870...12L, shappee2018seeing}. In this sample, SN~2011fe, SN~2015F and SN~2018gv are well-observed normal SNe~Ia; whereas, SN~2017cbv and SN~2018oh are among few SNe~Ia that have been detected with significant excess flux at early times ($\lesssim 5$ days after time of first light), and iPTF16abc has a nearly linear rise in $g$-band with relatively limited amount of data. SN 2018gv has photometric data in $g^\prime$ and $i^\prime$ band from Sinistro camera at the Las Cumbres Observatory
from very early phases. SN~2018oh has measurements in $g$ and $i$ from PanSTARRS within the first day after explosion, but due to the large uncertainty ($\gtrsim 0.3$ mag) we do not include them in the comparison. SN 2011fe lacks $i$-band data, but \cite{2011fe-specphot} has produced a spectrophotometric time series with high precision, which enables us to calculate synthetic photometry in the $g$, $i$ and \Kepler\ bands. SN~2017cbv has both photometric measurements and spectra in the early phase. The synthetic photometry of SN 2011fe and SN~2017cbv in $g$,$i$ and \Kepler\ bands are calculated through the same routine used in Sec.~\ref{calibration}.




 

\section{Discussion}

In the traditional `expanding fireball' model, the photospheric temperature and expanding velocity remains constant within days after explosion, and thus a SN~Ia light curve follows $L\propto t^2$ as ejecta expand in all bandpasses \citep{arnett1982type}. Observations suggests that light curves of different SNe~Ia may follow power-law $L\propto t^\alpha$ with some distribution of power law index $\alpha$ centered on $2$ \citep[e.g.][]{hayden2010rise, gaitan2012, 2020ApJ...902...47M}. The rise of SN~2018agk's light curve is well described by a single power-law with $\alpha = 2.25\pm 0.07$ (see Table \ref{tab:fitresult}), in good agreement with the other SNe~Ia from the \kepler\ prime mission \citep[][]{olling2015} and the ground-based samples. With the excellent cadence and high S/N of SN~2018agk's \kepler\ light curve, we can put strong constraints on the existence of excess flux, and we can also determine the time of the first light to a very high precision. Combined with the 1-day cadence DECam observations within the first 4 days after first light, we find that that there is no significant color evolution in the first $\sim 10$ days after explosion.

Clues to the progenitor systems can be found through early observations of SN~Ia light curves soon after explosion. Different mechanisms have been proposed to create early excess features with different characteristics in SN~Ia light curves, e.g. \cite{kasen2009seeing} 
shows that in the SD model, the interaction between SN ejecta and a non-degenerate companion can produce a blue excess on top of normal SN~Ia flux in light curves within days after explosion. Such early bump features have only been detected in a very limited number of normal SNe~Ia. For both SN~2017cbv and SN~2018oh, interaction with a subgiant companion can reproduce the early optical light curve within the uncertainty range, but the late-time spectra lack H and He features predicted by this model \citep{hosseinzadeh2017early, nebularspec18oh}. \cite{Levanon+19} further compared the three processes for the early light excess of SN~2018oh and argue that ejecta-DOM interaction in the DD scenario accounts for the early excess better than the companion-interaction or the presence of $^{56}\rm Ni$ in the outer ejecta. Nonetheless, the uncertainty in the underlying supernova flux in different bands at the earliest stage \redpen{strongly degrades the information that can be interpreted from the model fitting}. Early multi-band observations on SNe Ia with clearly no early excess features are crucial for future studies.

\subsection{Color Evolution of SNe Ia with and without Early Excess Flux}

The color evolution at the earliest phase of SNe~Ia can be a crucial piece of evidence in distinguishing different progenitor models, as discussed above. Still, the small sample of high quality measurements in this critical time window limits the application of such studies. Currently, there are not enough SNe~Ia light curves with sufficient cadence and S/N at these early epochs to build a reliable SN~Ia template\footnote{There do exist templates at the earliest stage, e.g. SiFTO \citep{2008ApJ...681..482C}, though the uncertainties are very large due to the small number of objects.}. 





To reveal the prototypical color evolution of normal SNe~Ia without excess flux and distinguish them from ones with excess flux, we show the color evolution of SNe~Ia with early observations in Figure \ref{fig:color evolution}. In general, the normal SNe~Ia without early excess (SN~2011fe, SN~2015F, SN~2018gv and SN~2018agk) in our sample have similar colors, with only a small color evolution toward the blue. In $g-i$, the color changes  less than $\sim 0.2$ mag in the first 10 days. Remarkably, there is a clear distinction between normal SNe~Ia and SN~2017cbv in $g-i$, where the latter is significantly blue not only during the time of excess, but up to 10 days after first light. In addition, the synthetic colors of SN~2017cbv ($g-$\kepler\ and \kepler$-i$ ) show a similar trend than $g-i$, they are bluer than SN~2011fe. On the other hand, iPTF16abc (with a linear rise in $g$-band) seems to have a similar $g-i$ color curve as normal SNe~Ia without excess. However, iPTF16abc shows very different behaviour in $B-V$ color \citep[][]{Stritzinger2018} and will be discussed later in more detail.
Notably, in \Kepler$-i$, the earliest high S/N photometric measurement of SN~2018oh at $t\sim1$ day after first light is significantly bluer than those of SN~2018agk, while the synthetic color of SN~2011fe and SN~2017cbv (empty symbols in Fig. \ref{fig:color evolution}) have smaller difference in the earliest stage in this band.
\begin{figure*}[]
\centering
\includegraphics[width=0.48\textwidth]{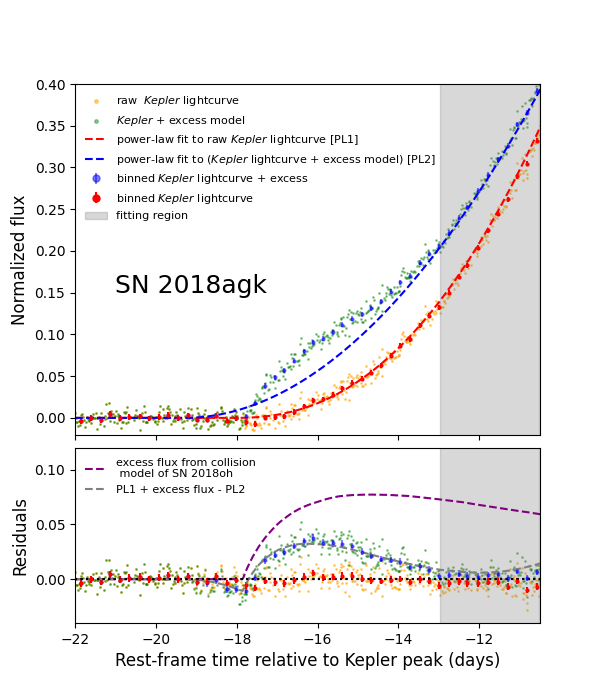}
\includegraphics[width=0.48\textwidth]{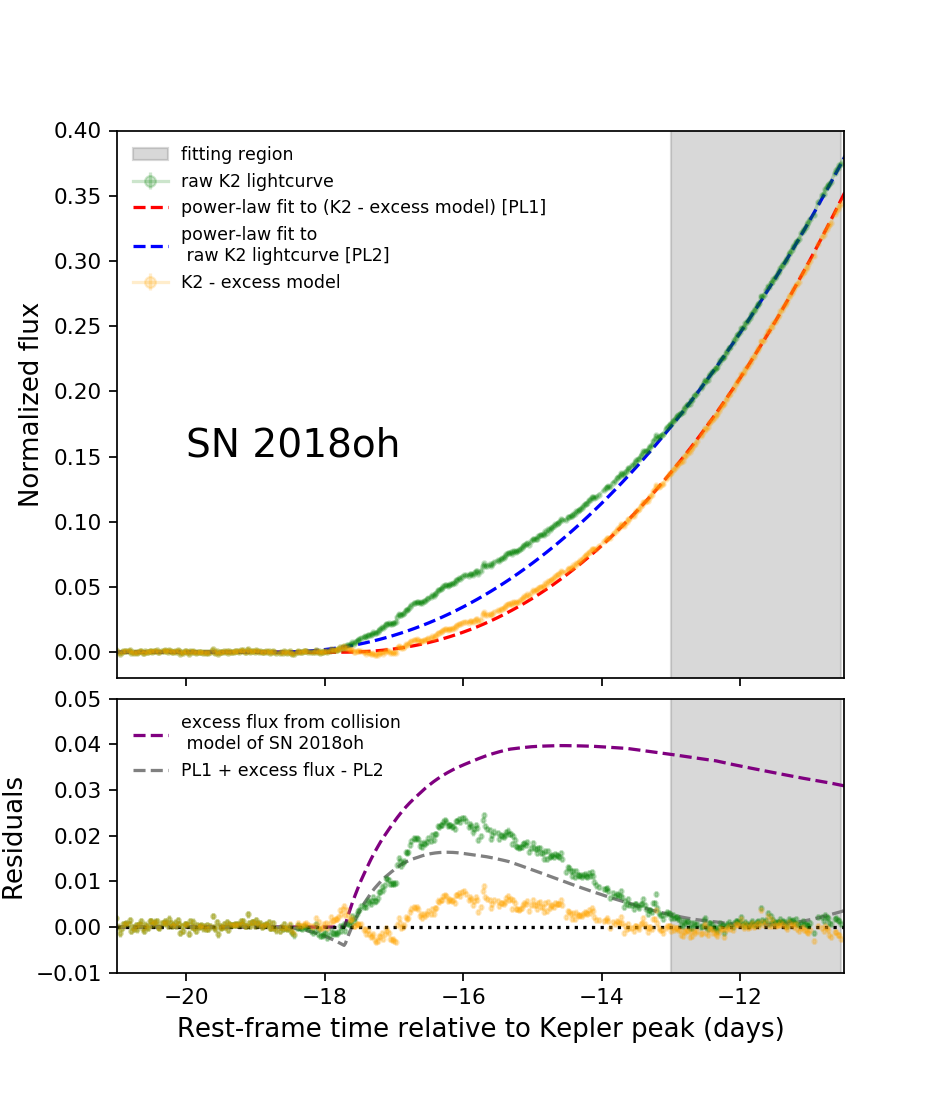}
\caption{\textbf{Left:} The top panel shows power-law fit to the original  \Kepler\ light curve of SN~2018agk (PL1, red) and \Kepler\ light curve plus long-tail excess from the companion-interaction model (PL2, blue). The bottom panel shows the residuals relative to each power-law fit. The excess model (purple) is also included for comparison in the bottom panel. The grey line represents the difference between PL1 plus the excess model and PL2. With the traditional power-law fit method for excess detection, the excess can be strongly underestimated, as can be seen from the comparison between the purple and grey dashed lines. The excess model comes from the fit to the bump of SN~2018oh and is added to SN~2018agk in physical units before normalized to the peak of SN~2018agk. We note that the small dip at $\sim-18$ day comes from the difference of $t_0$ of two power-law fittings. \textbf{Right:} same comparison between raw light curve of SN~2018oh in \Kepler\ (green and blue) and the light curve with excess model subtracted (yellow and red).
}
\label{fig:excess_fit}
\end{figure*}

This result is slightly different from previous studies in other optical bands. 
Previous statistical study on the ZTF SNe~Ia sample in $g$ and $i$ bands found relative homogeneity in early color ($g-i \sim 0.15$ mag), along with a large scatter of color slope in $g-r$ for individual events \citep{2020ApJ...902...47M,2020ztf3}.
\cite{Stritzinger2018} has analyzed a sample of 13 SNe~Ia with early measurements in $B$ and $V$-band, and find two populations with different color evolution in $B-V$ within the first $\sim 3$ days after explosion \redpen{(see Figure 2 in \cite{Stritzinger2018})}. The `early red' group has redder initial color and turns blue rapidly in $B-V$, with a rate of $\sim 0.5$ magnitude in the first 10 days and has typical luminosity and decline rate among SNe Ia. The `early blue' group is $\sim 0.5$ mag bluer than the `early red' group in $B-V$, evolves at a negligible rate and tends to be brighter than the red group. Spectroscopically, the SNe Ia in `early blue' group tend to have weaker Si II absorption features and lie within or close to the shallow silicon (SS) sub-type in Branch diagram (see Fig.\ref{fig:spec_config}), while the SNe~Ia in the `early red' group belong to either core normal (CN) or cool (CL) type. 
Qualitatively speaking, the spectral features of SN~2018agk fit into the core normal (CN) class (see Sec.~\ref{sec:spec_analysis}), and it has intermediate peak brightness and decline rate among the normal SNe~Ia sample. Combining all the characteristics, SN~2018agk aligns with the `early-red' events as characterized in \cite{Stritzinger2018}. 
In terms of early color, in $g-i$ the SNe~Ia in the `early-red' group (SN~2011fe, SN~2015F and SN~2018agk) are similar to iPTF16abc but are significantly redder than SN~2017cbv, both of which belong to `early blue' group. Meanwhile, all the SNe~Ia in our sample evolve with a slow rate no more than $\sim 0.02$ mag per day in the first 10 days in $g-i$, and the rapid drop of the `early-red' group in $B-V$ is not seen. 

\cite{2009ig} analyzed SN~2009ig, a SN~Ia discovered hours after explosion with significant $B-V$ color evolution at early times. The redder colors (e.g., $V-R$ and $R-I$) of SN~2009ig do not evolve quickly, indicating that the rapid $B-V$ color change is not caused by a broad wavelength continuum change.  Instead, the early spectra of SN~2009ig during this period show \ion{Si}{2} 4000 feature that is so blueshifted and broad that it merges with the Ca H\&K absorption feature.  As a result, the flux at 4000~\AA\ is severely depressed.  Over the next few days, corresponding to the time of rapid $B-V$ color evolution, the features become distinct and more similar to the appearance of SN~Ia spectra near peak brightness.  This strongly indicates that spectral features likely drive these color changes.  Such a mechanism may explain the difference in color evolution in different bands seen for the full sample, although a larger and more complete sample of SNe~Ia will be necessary to resolve this issue.

In general different models give qualitatively different prediction on the early color of excess in SNe~Ia.
Overall the comparison between normal SNe~Ia sample with SN~2017cbv agrees with fits from the companion-interaction model in \cite{hosseinzadeh2017early} and simulations of the $^{56}$Ni shell model from \cite{2020A&A...642A.189M}, both of which can produce a bluer color than the fiducial model without bump. Still, the exact prediction may vary with different model parameters, e.g. the fit to SN~2018oh has a redder color between $\sim3$-$8$ days after explosion compared to that of SN~2017cbv in \cite{magee2021exploring}.  Uncertainties in the fiducial models at the early time also limits our ability to fit models with high fidelity \citep[e.g. see the discussion in ][]{magee2021exploring}.

The $t_{fl}$ of SNe Ia in our sample likely face systematic uncertainies. They are inferred based on the $t^{\alpha}$ fit with slightly different schemes in separate papers, with varying lengths of the dark phase. Nonetheless, as shown in the Fig~\ref{fig:color evolution}, SNe Ia in our sample only have a very low level of variability in color at the early time. Therefore, the uncertainty in $t_{fl}$ should not have big influence on our conclusion. 



\begin{figure*}[t!]
\begin{center}
\scalebox{1.}
{\includegraphics[width=0.9\textwidth]{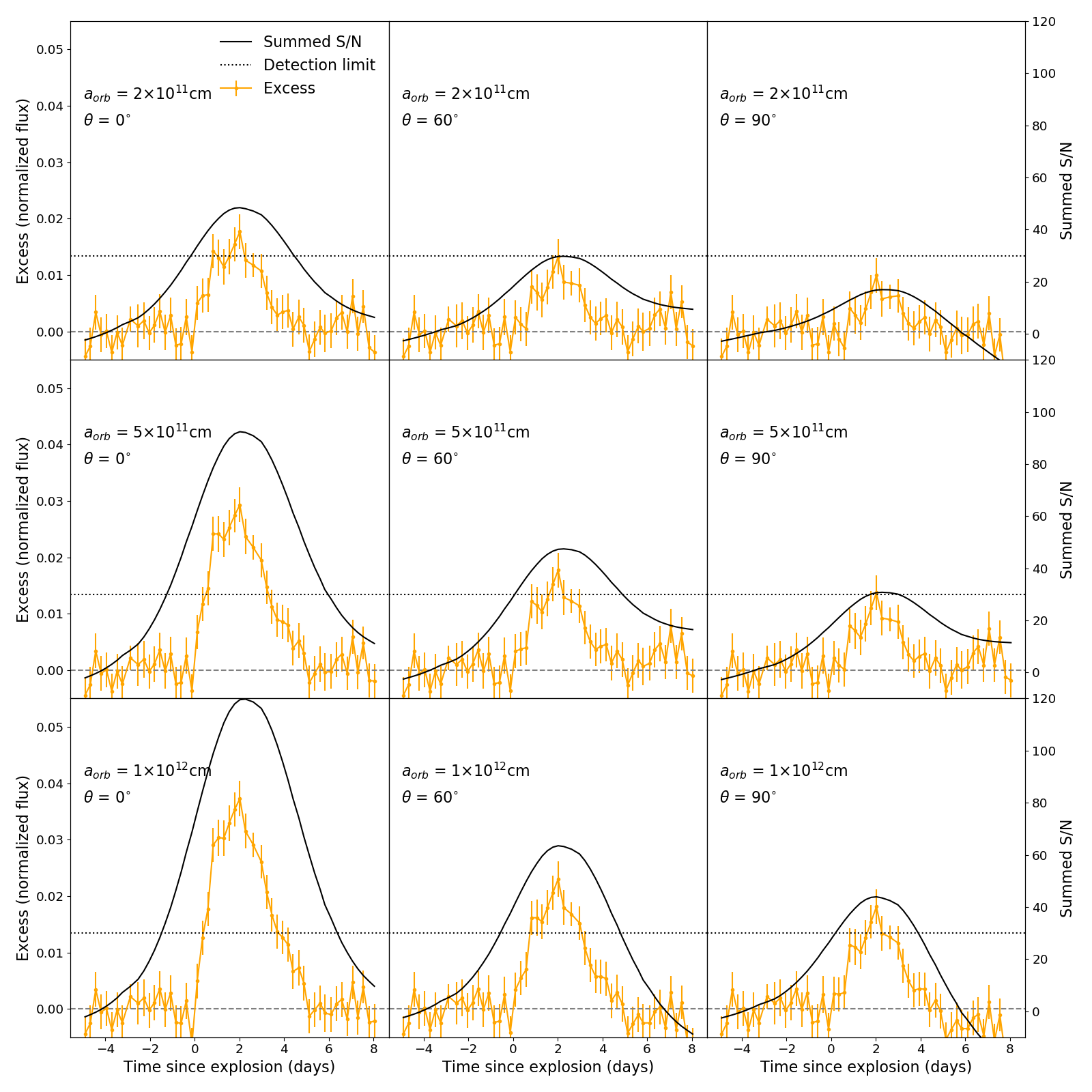}}
\caption{The excess relative to power-law fit of the light curve from companion-interaction model with different parameters, calculated in the same way as shown in Figure~\ref{fig:excess_fit}. The \kepler\ light curve of SN~2018agk is used to approximate the supernova flux in the model. The power-law is fitted on the time interval starting from $6.5$ day to $8$ day after explosion. The black curves are the rolling sum of S/N with Gaussian window. We conservatively set the the detection limit of excess as S/N$=30$, marked as the black horizontal dotted line. }
\label{fig:gassiansumexample} 
\end{center}
\end{figure*}

\subsection{Detectability of a Prominent Bump in SN\,2018agk}\label{sec:bumpdetection}

High cadence surveys including \kepler\ and \tess\ probe a large sample of SNe~Ia soon after their explosion, making it possible to study the early light curve of SNe~Ia statistically and estimate the fraction of SNe~Ia with detectable early excess features. Resolving the detectability of the early excess and potential contamination will be a key component to such studies.
Considering the current limitation in fitting with physical models, model-independent excess detection algorithms are necessary for searching abnormal features in the rise of SNe~Ia. In this section we present the application of the rolling sum algorithm to evaluate excess flux in early SNe~Ia light curves.

With the high quality light curve of SN~2018agk, we try to evaluate the factors influencing efficiency and accuracy of bump-detection quantitatively. We add the excess model onto SN~2018agk's light curve and evaluate the significance of bump detection with method described in \cite{dimitriadis2018k2}, which fits power-law to a time interval well-after the explosion as the baseline light curve and find the excess signal in the residual within days after explosion. We use an excess flux model based on a companion-interaction model fitted to SN~2018oh in \cite{dimitriadis2018k2}, with a subgiant companion at a separation of $2\times10^{12}$cm assuming the supernova ejecta has mass of $1.4M_\odot$ and velocity of $8\times10^8$cm s$^{-1}$. With the assumption that the luminosity of the bump has a weak dependence of supernova luminosity, we add the excess in physical units before normalizing to the peak of SN~2018agk. 

To minimize the degeneracy issue and induced high uncertainty in power-law fitting, we perform the Markov Chain Monte Carlo (MCMC) fitting with the \texttt{emcee} package \citep{foreman2013emcee} assuming a Gaussian prior distribution for $t_0$ and $n$ and a flat prior distribution for scale factor $A$. The mean and standard deviation of $t_0$ are set to be $18.7\pm1.8$ as estimated from \cite{2020ApJ...902...47M}, and we use power-law index in $r$-band $n_r = 2.0\pm0.5$ to simulate the distribution of $n$ in \Kepler\ bandpass, considering similar effective wavelength of \Kepler\ and $r$-band. 

The left panel of Fig.\ref{fig:excess_fit} shows the power-law fit to \Kepler\ light curve of SN~2018agk (PL1) and the same light curve plus the excess flux model (PL2), along with the residuals of these two power-law fitting. In the right panel of Fig.\ref{fig:excess_fit} we show the fitting with the same setting on SN~2018oh, with or without excess flux model subtracted. All of the fits are done in the window of $13$ to $10.5$~days before $t^{peak}_{K2}$. One noticeable difference is that in both fits $t_0$ inferred from light curves with excess is $\sim 1.5$ days earlier than $t_0$ inferred from original light curve. In the residual plot, we show the contrast between the excess model (purple dashed line) and the reconstructed bump after power-law fitting. Clearly, due to the extended tail of collision flux, the power-law fit overestimates the supernova flux and underestimates the excess flux by a large amount after subtraction. This is an indication of the low effectiveness of current excess detection routines. This problem will be exacerbated for ground-based surveys with much larger cadences or similar space telescopes with shallower detection limits (e.g. \textit{TESS}). Therefore, establishing a robust fiducial model for early light curves of SNe~Ia without excess is necessary for future search on early excess. Meanwhile, signatures in either early spectra or color evolution, as analyzed previously, would also be key evidence to facilitate bump detection and distinguish different models. 

\subsection{Constraints on Progenitor System in the Case of the SD Model}

We adapt the fitting scheme from Section~\ref{sec:bumpdetection} to give a qualitative estimate on detectability of the bump and constrain the potential progenitor system for SN~2018agk in the case of SD model. We use the analytic companion interaction model \citep{kasen2009seeing} multiplied with an angular dependency term calculated in \cite{brown2012ang} as a test case. We first add excess flux to its \kepler\ flux, and then fit a power-law to relatively late time interval with MCMC and estimate the detectability of excess in residuals within $\sim 5$ days after explosion. 
The excess flux is mainly determined by 4 parameters: the orbital separation of the binary $a_{orb}$; viewing angle $\theta$; the velocity $v$; and mass $m_{ejecta}$ of the ejecta. While $\alpha$ and $\theta$ can vary with relatively large ranges, $v$ and $m_{ejecta}$ are usually well constrained for typical SNe~Ia in the SD model. We set $v=8\times10^8$cm/s and $m_{ejecta} = 1.4M_\odot$ in accordance with a typical SNe~Ia explosion with kinetic energy of $10^{51}$ erg, and we focus on the effect of different viewing angle and orbital separation. 

\begin{figure}[t!]
\begin{center}
\scalebox{1.}
{\includegraphics[width=0.48\textwidth]{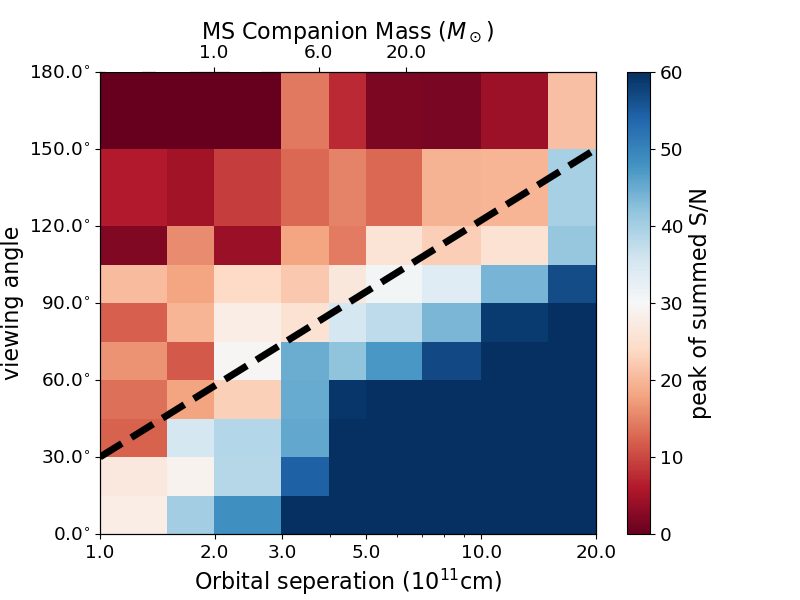}}
\caption{The peak of rolling summed S/N with Gaussian window for companion-interaction model with different orbital separations $a_{orb}$ and viewing angles $\theta$, as introduce in Figure~\ref{fig:gassiansumexample}. We empirically set the detection limit to be S/N $\approx30$, corresponding to the white blocks in the color map, and blue region represents the detectable excess, while in red region the excess is non-detectable. The black dash line qualitatively marks the bounds of detectability of excess in parameter space. The mass of a main sequence companion in the MS-WD binary system undergoing Roche lobe filling with certain orbital seperation is labeled at top.
}
\label{fig:bumpdetection} 
\end{center}
\end{figure}

To evaluate the overall signal of excess on top of the power-law rise, we applied the rolling method with Gaussian window to sum signal-to-noise of the `detected excess' over its extended duration. 
Considering that the typical duration of the detected bump is $\sim 5$~days, we use a Gaussian with $\sigma = 2$ days over a $12$-day window, and then we normalize the results to the rolling sum of a constant background to remove the edge effect in the algorithm. The left column of Fig. \ref{fig:gassiansumexample} shows a set of examples with $\theta=0^{\circ}$ and different orbital separation $a_{orb}$ and the left column shows results for fixed $a_{orb} = 3\times 10^{11}$cm at different viewing angles. As can be seen from the plot, this method can effectively take the accumulated signal in an extended interval into account and smooth out the random noise in the light curve at the same time. We conservatively set the detection limit as S/N $\approx30$ as the dotted line in Fig. \ref{fig:gassiansumexample}, which is well above the rolling sum of baseline S/N without excess, i.e. the model with viewing angle $\theta = 180^{\circ}$.  We run the same algorithm on excess for a large set of different $\theta$ and $a$ and plot the color map of the peak of summed S/N with regard to these two parameters in Fig. \ref{fig:bumpdetection}. The blue and red region represents the regions of detectable and non-detectable excess in the parameter space, and we add a black dashed line marking approximate detection limit for the SD progenitor system of SN~2018agk in Fig. \ref{fig:bumpdetection}

We are able to give a rough constraint on the potential progenitor system of SN~2018agk in the SD model and the detectability of excess for similar events with high cadence light curves. An assumption of the companion interaction model is that the companion overfills the Roche lobe, and the orbital separation $a$ is correlated with the companion radius $R$ and the mass ratio of the binary $q$, approximately following the equation \citep{1983ApJ...268..368E}:
\begin{equation}
	a_{orb} = \frac{0.6q^{2/3}+\ln(1+q^{1/3})}{0.49q^{2/3}}R.
\end{equation}
For simplicity we estimate a typical MS star in the WD-MS binary with the mass ranging from $1-20 M_\odot$ and mark the mass corresponding to certain $a_{orb}$ at the top of Fig. \ref{fig:bumpdetection}, though such binary systems are believed to likely produce less energetic explosions, e.g. nova \citep[][]{Prialnik1986,shara1986, Michaely2021}. In the more favorable SD models, the companions are usually believed to be sub-giants or red giants \citep[][]{1995PASP..107.1019B, hachisu1996, wang2010}. Such binary system will have larger $a_{orb}$ for the same mass range, and thus will only be more constrained compared to the WD-MS system analyzed here.
As shown in the Fig. \ref{fig:bumpdetection}, a WD-MS binary progenitor system with viewing angle $\theta\lesssim 60^{\circ}$ will produce a prominent excess that is detectable to our rolling algorithm for a typical SN~Ia like SN~2018agk in \kepler. When $\theta$ increase to $\gtrsim 120^{\circ}$, the excess becomes undetectable for such WD-MS binary systems. 
From this perspective, viewing angle seems to be the more dominant factor in the detectability of the bump, and a statistical analysis of SNe~Ia with high cadence early light curves from \kepler\ and \tess\ will be the key to determine if SD system can be a major channel for SNe~Ia or not.

\section{Conclusions}
In this paper, we report and analyze the photometric and spectroscopic observations of SN~2018agk, a normal SNe~Ia that occured within the \kepler\ Campaign 17 field. The SN has an exquisite high cadence \kepler\ light curve, showing no early excess flux beyond a power-law rise. Combined with ground-based photometry in multiple bands, especially the 1-day cadence DECam observations within the first $\sim 4$ days after first light in \kepler, we are able to determine the color evolution of a prototypical SNe~Ia without excess flux in the earliest stages. 
\begin{itemize}
\item The spectra of SN~2018agk match the features of a branch normal SNe~Ia without peculiarity, e.g. the pEW and velocity of Si II fit in the range of normal SNe~Ia. Especially, the \textit{HST} UV spectrum near peak shows similarities in continuum levels and line features with SN~2011fe, indicating similar progenitor metallicity of the two SNe~Ia.
\item The early \kepler\ light curve shows a clear power-law rise. Using a BIC test, we find that a single power-law with an index of $\alpha = 2.25\pm0.07$ fits the data better than a canonical fireball model ($L\propto t^2$) or a double power-law fit. There is no evidence of early excess flux in the fit. 
\item Combined with DECam-$g$ and $i$ observations in the first $\sim 4$ days after first light in \textit{Kepler}, we calculate the early color evolution and compare it with other SNe~Ia with and without early excess flux. We show that the normal SNe~Ia without bumps have similar color in the first $\sim 10$ days after first light. Of the SNe~Ia with bumps, SN~2017cbv has bluer color during this period, especially in $g-i$, but SN~2015F has similar color as other normal SNe~Ia. All SNe~Ia in our sample have very low levels of variability in color during this early rise phase ($|\Delta(g-i)/\Delta t| \lesssim 0.02$ mag day$^{-1}$), in contrast to the steep evolution in $g-r$ for a large portion of SNe~Ia in \cite{2020ztf3}, and the rapid drop in $B-V$ for some `early red' SNe~Ia as found in \cite{Stritzinger2018}. Qualitatively speaking, this result agrees with the predictions by both the companion interaction model and the $^{56}Ni$ shell model, but a larger sample and simulations with higher precision are necessary for further differentiate between these progenitor channels.
\item We examine the efficiency in bump detection by adding the excess model derived from a fit to SN~2018oh to the \kepler\ light curve of SN~2018agk, and applying the detection algorithm from \cite{dimitriadis2018k2}. With an extended tail into $\gtrsim 5$ days after explosion, the excess can strongly influence the power-law fitting results, shifting $t_{fl}$ by $\sim 1$ day even in fitting with strong prior distribution, and the `bump' can be significantly underestimated. Thus, a thorough analysis on the efficiency and contamination of bump detection will be needed for future studies.
\item We further test the detectability of bumps in the companion interaction model and constrain the physical properties of possible SD progenitor with the rolling sum method. 
We found that in the case of SN~2018agk, a MS or more evolved companion (e.g. sub-giant or red giant) with mass larger than $1 M_\odot$ and viewing angle smaller than $60^{\circ}$ can be ruled out, while the excess from WD-MS system with viewing angle larger than $120^{\circ}$ is still below the detection limit. On the other hand, the influence of MS companion mass on the detectability of excess is relatively low. A statistical analysis of SNe~Ia sample from \kepler\ and \tess\ will be the key to revealing if SD progenitor systems are the main channel for SNe~Ia or not.
\end{itemize}

Overall, SN~2018agk has no signatures of spectroscopic and photometric peculiarities. Its early \kepler\ light curve closely follows a power-law rise and has no signature of excess flux, and early DECam observations show no early color evolution of this event. Thus, SN~2018agk can serve as a prototype for normal SNe~Ia without early excess for future studies on searching for early excess. 

While it is difficult to exclude any progenitor system for SN~2018agk alone, a statistical study on the SNe~Ia sample with high cadence light curve from \kepler\ and \tess\ will be a brand new window to shed light on the progenitor system of SNe~Ia. To facilitate such investigation, a complete statistical study on the detectability of early excess in different models will be crucial.

\

\textbf{Acknowledgments:}

KEGS is supported in part by NASA K2 cycle 4, 5, and 6 grants NNX17AI64G, 80NSSC18K0302, and 80NSSC19K0112, respectively. \redpen{Some of the data presented in this paper were obtained from the Mikulski Archive for Space Telescopes (MAST) at the Space Telescope Science Institute. The specific observations analyzed can be accessed via \dataset[10.17909/T93W28]{https://doi.org/10.17909/T93W28} and \dataset[10.17909/T97P46]{https://doi.org/110.17909/T97P46}.}

This research is based on observations at Cerro Tololo Inter-American Observatory, National Optical Astronomy Observatory (NOAO 2017B-0279; PI: A Rest, NOAO 2017B-0285; PI: A Rest), which is operated by the Association of Universities for Research in Astronomy (AURA) under a cooperative agreement with the National Science Foundation.

This research is based on observations obtained at the international Gemini Observatory, which is managed by the Association of Universities for Research in Astronomy (AURA) under a cooperative agreement with the National Science Foundation on behalf of the Gemini Observatory partnership: the National Science Foundation (United States), National Research Council (Canada), Agencia Nacional de Investigaci\'{o}n y Desarrollo (Chile), Ministerio de Ciencia, Tecnolog\'{i}a e Innovaci\'{o}n (Argentina), Minist\'{e}rio da Ci$\hat{\text{e}}$ncia, Tecnologia, Inova\c{c}\~{o}es e Comunica\c{c}\~{o}es (Brazil), and Korea Astronomy and Space Science Institute (Republic of Korea). Observations in this program were obtained through program ID GS-2017B-LP-13.

The UCSC transient team is supported in part by NASA/{\it K2} grants 80NSSC18K0303 and 80NSSC19K0113, the Gordon \& Betty Moore Foundation, the Heising-Simons Foundation, and by a fellowship from the David and Lucile Packard Foundation to R.J.F.  D. O. J. acknowledges support provided by NASA Hubble Fellowship grant HST-HF2-51462.001, which is awarded by the Space Telescope Science Institute, operated by the Association of Universities for Research in Astronomy, Inc., for NASA, under contract NAS5-26555.

J.V. and the Konkoly team has been supported by the project "Transient Astrophysical Objects"  GINOP 2.3.2-15-2016-00033  of the National Research, Development and Innovation Office (NKFIH), Hungary, funded by the European Union.

The LCOGT team is supported by NASA grant 80NSSC19
K0119 and NSF grants AST-1911225 and AST-1911151.

Pan-STARRS is a project of the Institute for Astronomy of the University of Hawai'i, and is supported by the NASA SSO Near Earth Observation Program under grants 80NSSC18K0971, NNX14AM74G, NNX12AR65G, NNX13AQ47G, NNX08AR22G, and by the State of Hawai'i. 

This paper uses data obtained with ANDICAM mounted to the 1.3m telescope at the Cerro Tololo Inter-American Observatory (CTIO) and operated by the SMARTS Consortium under programme NOAO-18A-0047 (PI: Galbany).

QUB acknowledges funding from STFC Grants ST/S006109/1, ST/P000312/1 and ST/T000198/1.

This project has been supported by the Lend\"ulet Program of the Hungarian Academy of Sciences, project No. LP2018-7/2020.
Research infrastructure was provided by the Hungarian Academy of Sciences.

This work was partially supported by the Center for Astrophysical Surveys (CAPS) at the National Center for Supercomputing Applications (NCSA), University of Illinois Urbana-Champaign.

Parts of this research were supported by the Australian Research Council Centre of Excellence for All Sky Astrophysics in 3 Dimensions (ASTRO 3D), through project CE170100013.

This research has made use of the SVO Filter Profile Service (http://svo2.cab.inta-csic.es/theory/fps/) supported from the Spanish MINECO through grant AYA2017-84089


Q.W. acknowledges financial support provided by the STScI Director's Discretionary Fund.

Y.Z. thanks Alexey Bobrick and Naveh Levanon for valuable discussions.

M.R.M. is funded by the EU H2020 ERC grant no. 758638.

D.A.C. acknowledges support from the National Science Foundation Graduate Research Fellowship under Grant DGE1339067.

L.G. acknowledges financial support from the Spanish Ministry of Science, Innovation and Universities (MICIU) under the 2019 Ram\'on y Cajal program RYC2019-027683 and from the Spanish MICIU project PID2020-115253GA-I00.

D.O.J. acknowledges support provided by NASA Hubble Fellowship grant HST-HF2-51462.001, which is awarded by the Space Telescope Science Institute, operated by the Association of Universities for Research in Astronomy, Inc., for NASA, under contract NAS5-26555.

PG and PC acknowledge the support of NASA grant NAS5-26555 from program HST GO-15274. We also thank R. Kirshner for his support with this work.

L.W. is supported by NSF grant AST-1817099 and NASA grant and NASA grant 80NSSC20K0538. 

L.K. acknowledges the financial support of the Hungarian National Research, Development and Innovation Office grant NKFIH PD-134784. L.K is a Bolyai János Research Fellow.

M.G. is supported by the EU Horizon 2020 research and innovation programme under grant agreement No 101004719.

J.B. would like to thank Lisa Rush, Piper English and Andre Van Zundert for their help in data collection.

S.G.G. acknowledges support  by FCT  under  Project  CRISP  PTDC/FIS-AST-31546/2017  and UIDB/00099/2020.

J.R.S. is funded by FCT (PD/BD/150487/2019), via the IDPASC PhD program, and by the CRISP project (PTDC/FIS-AST/31546/2017).

B.E.T. acknowledge parts of this research was carried out on the traditional lands of the Ngunnawal people. We pay our respects to their elders past, present, and emerging. B. E. T. and his group were supported by the Australian Research Council Centre of Excellence for All Sky Astrophysics in 3 Dimensions (ASTRO 3D), through project number CE170100013.

M.~R.~S. is supported by the NSF Graduate Research Fellowship Program Under grant 1842400.

S.W.J. acknowledges support from US National Science Foundation award AST-1615455.

J.B. is supported by NSF grants AST-1313484 and AST-1911225, as well as by NASA grant 80NSSC19kf1639.

M.N. is supported by a Royal Astronomical Society Research Fellowship and by the European Research Council (ERC) under the European Union’s Horizon 2020 research and innovation programme (grant agreement No.~948381).

The work of X.W. has been provided by the National Science Foundation of China (NSFC grants 12033003 and 11633002), the Major State Basic Research Development Program (grant 2016YFA0400803), and the Scholar Program of Beijing Academy of Science and Technology (DZ:BS202002).
\\

\textit{Software:} Astropy\citep{astropy18}, IRAF \citep{iraf93}, SAOImage DS9 \citep{saods9}, sncsomo \citep{sncosmo}, Matplotlib \citep{matplotlib}, SciPy \citep[][]{2020SciPy-NMeth}, NumPy \citep{2020NumPy-Array}, emcee \citep{foreman2013emcee}, celerite \citep{celerite}, corner\citep{corner}, extinction\citep{extinction}, pysynphot \citep[][]{2013ascl.soft03023S}, 
swarp \citep[][]{terapix2002}
Dophot \citep{dophot},  
HOTPANTS \citep{HOTPANTS}.\\

\facilities{Kepler, HET, Gemini:Gillett, SSO:1m, Du Pont, NTT, FTN, Shane, Keck:I, SOAR, Blanco, Swope, CTIO:1.3m, PS1}


\bibliographystyle{yahapj}
\bibliography{references}





\section{Tables}

\startlongtable
\begin{deluxetable*}{ccccccc}
\centering
\tablecolumns{7} \tablewidth{3pc} 
\tablecaption{Log of Spectroscopic Observations of SN 2018agk}
\tablehead{\colhead{MJD} & \colhead{\tablenotemark{a}Phase [days]} & \colhead{Telescope} & \colhead{Instrument} & \colhead{Exposure time [{s}]} & \colhead{Grism/Grating} & \colhead{Wavelength Range [{\AA}]}}
\startdata 
58193.34 & -10.6 & HET & LRS2 & 2000 & Blue & 3640-6970\\
 58194.45 & -9.5 & Gemini-North & GMOS & 3$\times$900 & R400 & 3957-8688 \\ 
 58196.26 & -7.7  & du Pont & WFCCD & 1200 & Blue/Red & 3500-9500 \\
 58196.72 & -7.3 & HST & STIS & 2100 & G230L & 1570–3180 \\ 
58200.32 & -3.8 & HET & LRS2 & 1800 & Blue & 3640-6970\\
 58201.26 & -2.8  & NTT & EFOSC2 & 2$\times$900 & gr11+gr16 & 3340-10000 \\
 58201.69 & -2.4 & HST & STIS & 2100 & G230L & 1570–3180 \\ 
 58201.88 & -2.3 & HST & STIS & 800 & G430L & 2900–5700 \\ 
 58201.90 & -2.2 & HST & STIS & 600 & G750L & 5240–10270 \\ 
 58214.25 & +9.8   & NTT & EFOSC2 & 1500  & gr11 & 3340-7456 \\
 58214.52 & +10.1 & FTS & FLOYDS & 3600 & - & 3146-10868\\
 58216.31 & +11.8  & NTT & EFOSC2 & 1500 & gr16 & 6000-10000 \\
 58217.55 & +13.0 & FTS & FLOYDS & 3600 & - & 3146-10868\\
 58225.34 & +20.6 & FTN & FLOYDS & 3600 & - & 3146-10885\\
 58229.17 & +24.4  & NTT & EFOSC2 & 2$\times$1500 & gr11+gr16 & 3340-10000 \\
 58233.32 & +28.4  & Shane & Kast & 4$\times$1200 & 452/3306+300/7500 & 3306-10495 \\
 58234.46 & +29.5 & FTN & FLOYDS & 3600 & - & 3146-10885\\
 58248.24 & +42.9 & FTN & FLOYDS & 3600 & - & 3146-10885\\
 58248.41 & +43.1  & Keck I & LRIS & 2$\times$1200 & 600/4000+400/8500 & 3122-10147 \\
 58287.09 & +80.8  & SOAR & Goodman & 2$\times$1800 & 400 M1+M2 & 3655-8896 \\
 58310.30 & +103.4  & Keck I & LRIS & 2$\times$1200 & 600/4000+400/8500 & 3122-10147 
\enddata
\tablenotetext{a}{Phases relative to B-band maximum on MJD 58204.178 according to SALT2 fit.}
\tablenotetext{}{Note: Uncertainties are in units of 0.001 mag.}%
\label{tab:spec}
\end{deluxetable*}

\startlongtable
\centering
\begin{deluxetable*}{cccccccccc}
\tabletypesize{\footnotesize}
\tablecaption{Ground-based optical photometry of SN~2018agk}
\tablehead{\colhead{MJD} & \colhead{\tablenotemark{a}Phase [days]} & \colhead{B(mag)} & \colhead{V(mag) } & \colhead{R(mag)} & \colhead{I(mag)} & \colhead{g(mag)} & \colhead{r(mag)} & \colhead{i(mag)} & \colhead{Telescope/Observatory}}
\startdata 
58186.35 & -17.37 & - & - & - & - & - & - & 22.223(166) & Decam \\
58187.30 & -16.45 & - & - & - & - & 20.992(045) & - & - & Decam \\
58187.35 & -16.40 & - & - & - & - & - & - & 21.058(068) & Decam \\
58187.36 & -16.39 & - & - & - & - & - & - & 20.868(047) & Decam \\
58187.41 & -16.34 & - & - & - & - & - & - & 20.711(092) & Decam \\
58187.50 & -16.25 & - & - & - & - & 20.971(166) & 20.425(106) & - & PS1 \\
58187.56 & -16.19 & - & - & - & - & 20.683(232) & - & - & PS1 \\
58188.30 & -15.47 & - & - & - & - & 20.321(022) & - & - & Decam \\
58188.36 & -15.41 & - & - & - & - & - & - & 20.281(022) & Decam \\
58188.58 & -15.20 & - & - & - & - & 20.131(280) & - & - & PS1 \\
58193.12 & -10.78 & - & - & - & - & 18.062(006) & 18.014(007) & 18.068(010) & Swope \\
58193.15 & -10.75 & 18.237(039) & 18.006(048) & - & - & 18.182(034) & 18.021(042) & - & Las Cumbres \\
58193.17 & -10.73 & 18.174(014) & 18.014(017) & - & - & - & - & - & Swope \\
58193.37 & -10.53 & 18.045(041) & 17.989(052) & - & - & 18.088(034) & 17.768(044) & - & Las Cumbres \\
58193.40 & -10.50 & 18.165(015) & 17.997(016) & - & - & 17.988(012) & 17.935(014) & 17.987(019) & Swope \\
58194.23 & -9.69 & 17.996(025) & - & - & - & - & - & - & Las Cumbres \\
58194.24 & -9.68 & 17.970(026) & 17.907(029) & - & - & 17.884(024) & - & - & Las Cumbres \\
58194.25 & -9.67 & - & - & - & - & 17.887(024) & 17.831(026) & 17.812(046) & Las Cumbres \\
58194.26 & -9.66 & - & - & - & - & - & - & 17.793(046) & Las Cumbres \\
58194.30 & -9.63 & 17.904(010) & 17.766(011) & - & - & 17.728(008) & 17.693(009) & 17.758(012) & Swope \\
58194.50 & -9.43 & - & - & - & - & 17.822(014) & - & 17.772(013) & PS1 \\
58194.56 & -9.37 & - & - & - & - & 17.791(016) & - & - & PS1 \\
58195.26 & -8.69 & 17.752(010) & 17.585(010) & - & - & 17.609(007) & 17.513(008) & 17.650(010) & Swope \\
58195.40 & -8.55 & - & - & - & - & 17.551(003) & - & 17.668(002) & Decam \\
58195.45 & -8.50 & - & - & - & - & 17.668(012) & 17.578(010) & - & PS1 \\
58195.52 & -8.44 & - & - & - & - & 17.565(012) & - & - & PS1 \\
58196.21 & -7.76 & 17.577(008) & 17.415(009) & - & - & 17.425(006) & 17.332(007) & 17.452(009) & Swope \\
58196.40 & -7.58 & - & - & - & - & 17.338(002) & - & 17.503(002) & Decam \\
58196.45 & -7.53 & - & - & - & - & 17.511(026) & - & - & PS1 \\
58197.22 & -6.78 & 17.412(011) & 17.226(012) & - & - & 17.229(008) & 17.204(010) & 17.292(011) & Swope \\
58197.25 & -6.75 & 17.416(024) & 17.340(026) & - & - & 17.433(023) & 17.307(024) & 17.323(044) & Las Cumbres \\
58197.44 & -6.57 & - & - & - & - & 17.297(010) & 17.238(008) & - & PS1 \\
58197.54 & -6.47 & - & - & - & - & 17.317(010) & - & - & PS1 \\
58199.02 & -5.03 & 17.209(050) & 17.081(034) & 16.704(083) & 16.702(067) & - & - & - & Konkoly \\
58199.58 & -4.48 & - & - & - & - & 17.127(011) & - & - & PS1 \\
58200.08 & -3.99 & 17.097(031) & 17.271(040) & - & - & - & - & - & Las Cumbres \\
58200.09 & -3.98 & - & 17.046(037) & - & - & 17.126(031) & 17.174(035) & - & Las Cumbres \\
58200.10 & -3.97 & - & - & - & - & - & 17.065(032) & 17.019(055) & Las Cumbres \\
58200.20 & -3.88 & 17.133(010) & 17.010(012) & - & - & 17.005(007) & 16.915(009) & 17.147(011) & Swope \\
58201.25 & -2.85 & 17.066(011) & 16.917(011) & - & - & 16.907(010) & 16.744(022) & 17.094(011) & Swope \\
58202.05 & -2.07 & 17.083(035) & 16.942(032) & 16.625(043) & 16.770(058) & - & - & - & Konkoly \\
58202.31 & -1.82 & - & - & - & - & 16.937(007) & 16.859(008) & 17.205(012) & Swope \\
58203.24 & -0.91 & 17.063(011) & - & - & - & - & - & - & Swope \\
58203.26 & -0.89 & 17.087(015) & 16.900(014) & - & - & 16.936(008) & 16.809(011) & 17.224(014) & Swope \\
58203.30 & -0.85 & 17.085(043) & 16.886(036) & - & - & 17.016(036) & 16.869(038) & 17.063(062) & Las Cumbres \\
58204.20 & +0.02 & - & - & 16.755(095) & 16.710(100) & - & - & - & ANDICAM \\
58204.36 & +0.18 & 17.118(013) & 16.955(015) & - & - & 16.904(011) & 16.832(013) & 17.314(024) & Swope \\
58205.14 & +0.94 & 17.072(018) & 16.930(017) & - & - & 16.937(015) & 16.883(014) & 17.432(026) & Swope \\
58206.20 & +1.97 & 17.173(027) & 16.982(028) & - & - & 17.055(024) & 16.946(024) & 17.318(043) & Las Cumbres \\
58207.23 & +2.98 & - & - & 16.754(116) & 16.902(118) & - & - & - & ANDICAM \\
58210.92 & +6.57 & 17.535(122) & 17.186(068) & 16.887(058) & 17.238(084) & - & - & - & Konkoly \\
58211.21 & +6.85 & 17.397(030) & 17.101(030) & - & - & - & - & - & Las Cumbres \\
58211.22 & +6.86 & - & 17.075(029) & - & - & 17.221(025) & 17.049(027) & - & Las Cumbres \\
58211.23 & +6.87 & - & - & - & - & - & 17.091(027) & 17.524(049) & Las Cumbres \\
58211.23 & +6.88 & - & - & 16.995(113) & 17.085(080) & - & - & - & ANDICAM \\
58211.87 & +7.50 & 17.480(096) & 17.135(037) & 16.803(064) & 17.305(131) & - & - & - & Konkoly \\
58212.96 & +8.56 & 17.507(047) & 17.181(036) & 16.878(037) & 17.276(067) & - & - & - & Konkoly \\
58215.07 & +10.62 & - & - & - & - & 17.346(004) & - & - & Decam \\
58215.13 & +10.67 & - & - & - & - & 17.385(003) & - & - & Decam \\
58215.18 & +10.72 & - & - & - & - & - & - & 18.091(005) & Decam \\
58215.31 & +10.85 & 17.745(031) & 17.288(030) & - & - & 17.486(025) & - & - & Las Cumbres \\
58215.32 & +10.86 & - & - & - & - & 17.485(028) & 17.367(029) & 17.942(058) & Las Cumbres \\
58215.33 & +10.87 & - & - & - & - & - & - & 17.931(057) & Las Cumbres \\
58216.21 & +11.73 & 17.721(024) & 17.262(020) & - & - & 17.448(016) & 17.264(017) & 18.009(029) & Swope \\
58217.00 & +12.50 & 17.828(052) & 17.279(039) & 16.993(057) & 17.400(093) & - & - & - & Konkoly \\
58217.17 & +12.66 & - & - & 17.420(095) & 17.431(092) & - & - & - & ANDICAM \\
58217.25 & +12.74 & 17.888(019) & 17.377(015) & - & - & 17.571(014) & 17.359(015) & 18.018(026) & Swope \\
58217.92 & +13.39 & 17.832(052) & 17.294(037) & 17.070(052) & 17.436(074) & - & - & - & Konkoly \\
58218.90 & +14.35 & 18.166(048) & 17.530(036) & 17.384(052) & 17.617(084) & - & - & - & Konkoly \\
58219.63 & +15.06 & 18.206(029) & 17.623(029) & - & - & - & - & - & Las Cumbres \\
58219.64 & +15.07 & - & 17.568(030) & - & - & 17.867(027) & 17.589(030) & - & Las Cumbres \\
58219.65 & +15.08 & - & - & - & - & - & 17.633(029) & 18.088(056) & Las Cumbres \\
58219.89 & +15.31 & 18.162(051) & 17.547(040) & 17.303(062) & 17.709(085) & - & - & - & Konkoly \\
58220.42 & +15.83 & - & - & - & - & 17.801(015) & - & 18.120(017) & PS1 \\
58220.48 & +15.89 & - & - & - & - & 17.837(015) & - & - & PS1 \\
58220.89 & +16.29 & 18.377(073) & 17.645(052) & 17.345(072) & 17.425(115) & - & - & - & Konkoly \\
58221.28 & +16.67 & 18.355(021) & 17.619(014) & - & - & 17.925(015) & 17.524(013) & 18.042(020) & Swope \\
58221.41 & +16.79 & - & - & - & - & 17.920(015) & 17.594(011) & - & PS1 \\
58221.89 & +17.26 & 18.535(060) & 17.684(047) & 17.333(057) & 17.691(081) & - & - & - & Konkoly \\
58223.26 & +18.60 & 18.746(034) & 17.788(029) & - & - & 18.200(025) & 17.700(025) & 18.057(045) & Las Cumbres \\
58224.10 & +19.42 & 18.672(023) & 17.817(019) & - & - & 18.196(013) & 17.600(013) & 18.082(022) & Swope \\
58226.27 & +21.53 & 19.116(048) & 17.871(020) & - & - & 18.336(019) & 17.572(016) & 17.900(024) & Swope \\
58226.91 & +22.15 & 18.953(072) & 17.979(059) & 17.419(073) & 17.541(095) & - & - & - & Konkoly \\
58227.24 & +22.48 & 19.075(048) & 18.001(036) & - & - & 18.579(030) & - & - & Las Cumbres \\
58227.25 & +22.49 & - & - & - & - & 18.579(031) & 17.737(029) & 17.957(053) & Las Cumbres \\
58228.92 & +24.11 & 19.258(089) & 17.944(059) & 17.445(068) & 17.416(075) & - & - & - & Konkoly \\
58231.60 & +26.72 & - & 18.327(045) & - & - & - & - & - & Las Cumbres \\
58231.61 & +26.73 & - & 18.325(049) & - & - & 18.984(036) & 17.884(031) & - & Las Cumbres \\
58231.62 & +26.74 & - & - & - & - & - & 17.871(032) & 17.891(053) & Las Cumbres \\
58242.48 & +37.33 & - & - & - & - & - & 18.519(044) & - & Las Cumbres \\
58242.88 & +37.72 & - & 18.654(080) & 17.940(101) & 17.991(101) & - & - & - & Konkoly \\
58247.48 & +42.20 & 20.219(049) & 19.095(034) & - & - & 19.816(034) & 18.699(042) & 18.704(047) & Las Cumbres \\
58253.16 & +47.74 & 19.678(038) & 19.236(033) & - & - & - & - & - & Las Cumbres \\
58253.18 & +47.76 & - & 19.291(035) & - & - & 19.843(053) & 18.926(031) & 19.022(050) & Las Cumbres \\
58257.17 & +51.64 & - & - & - & - & - & 19.067(045) & - & Las Cumbres \\
58258.33 & +52.77 & - & - & - & - & 19.879(104) & 19.079(055) & - & PS1 \\
58258.40 & +52.84 & - & - & - & - & 19.779(080) & - & - & PS1 \\
58259.37 & +53.79 & - & - & - & - & 19.883(074) & - & 19.277(050) & PS1 \\
58260.34 & +54.73 & - & - & - & - & 19.728(059) & 19.150(031) & - & PS1 \\
58260.42 & +54.81 & - & - & - & - & 19.909(081) & - & - & PS1 \\
58261.00 & +55.38 & - & - & - & - & - & 19.259(048) & - & Las Cumbres \\
58272.15 & +66.24 & - & - & - & - & - & 19.488(045) & - & Las Cumbres \\
58272.16 & +66.25 & - & - & - & - & - & 19.619(050) & - & Las Cumbres 
\enddata
\tablenotetext{a}{Phases relative to B-band maximum on MJD 58204.178 according to SALT2 fit.}
\tablenotetext{}{Note: Uncertainties are in units of 0.001 mag. This table has also been published in its entirety in the machine-readable format, including individual measurements and upper-limits. A compact presentation is shown here for guidance regarding its form and content. Measurements with uncertainty greater than $0.3$ mag are not included in this table.}
\end{deluxetable*}

\startlongtable
\begin{deluxetable*}{cccccc}
\tablecaption{Ground-based infrared photometry of SN~2018agk}
\tablehead{\colhead{MJD} & \colhead{\tablenotemark{a}Phase [days]} & \colhead{H(mag)} & \colhead{J(mag)} & \colhead{K(mag)} & \colhead{Telescope}}
\startdata 
58204.20 & +0.02 & 17.220(252) & 16.937(107) & - & ANDICAM \\
58204.28 & +0.10 & 17.276(064) & 16.821(056) & 16.342(057)& SOFI \\
58207.23 & +2.98 & 17.694(189) & 17.313(237) & - & ANDICAM \\
58211.23 & +6.88 & 17.545(200) & 17.915(273) & - & ANDICAM \\
58216.11 & +11.62 & 17.590(064) & 17.813(081) & 16.765(070)& SOFI \\
58217.17 & +12.66 & 17.647(291) & 17.763(063) & - & ANDICAM \\
58231.17 & +26.31 & 17.439(066) & 17.687(081) & 16.830(085)& SOFI \\
58251.02 & +45.65 & 18.078(085) & 18.174(099) & 17.587(125)& SOFI 
\enddata
\tablenotetext{a}{Phases relative to B-band maximum on MJD 58204.178 according to SALT2 fit.}
\tablenotetext{}{Note: Uncertainties are in units of 0.001 mag.}
\end{deluxetable*}

\end{document}